\documentclass[prd,aps,12pt,nofootinbib,10pt,showpacs,showkeys]{revtex4}
\usepackage{graphicx}
\usepackage{epsfig}
\usepackage{bm}
\usepackage{amssymb}

\newcommand{\ham}{\mathcal H}
\newcommand{\sch}{Schr$\ddot{\mbox{o}}$dinger}

\newcommand{\lag}{\mathcal L}
\newcommand{\gcon}{\mathcal G}
\newcommand{\hh}{\mathcal H}
\newcommand{\ff}{\mathcal F}

\newcommand{\nc}{\newcommand}       
\nc{\nuc}[2]    {$^{#1}${#2}}           
\nc{\vc}[1] {\mbox{\boldmath $#1$}} 
\nc{\wtil}      {\widetilde}            
\nc{\del}       {\partial}              
\nc{\al}        {\alpha}                
\nc{\beq}     {\begin{eqnarray}}
\nc{\eeq}    {\end{eqnarray}}
\nc{\bra}       {\langle}               
\nc{\ket}       {\rangle}               
\nc{\bras}[1]   {\langle #1|}           
\nc{\kets}[1]   {|#1\rangle}            
\nc{\fra}     {\frac{1}{2}}
\nc{\nn}      {\\ \nonumber}
\nc{\mapleft}[1]{           
 \smash{\mathop{\,          %
  \hbox to 1.5cm{\rightarrowfill}\, }\limits_{#1}}}

\begin{document}
\title{Pseudoscalar Mesons in the $SU(3)$ Linear Sigma
Model \\ with Gaussian Functional Approximation}
\author{Hua-Xing Chen$^{1,3}$} \email{hxchen@rcnp.osaka-u.ac.jp}

\author{V. Dmitra\v sinovi\' c$^2$} \email{dmitrasin@yahoo.com}

\author{Hiroshi Toki$^1$} \email{toki@rcnp.osaka-u.ac.jp}

\affiliation{ $^1$Research Center for Nuclear Physics, Osaka
University 10-1 Mihogaoka, Ibaraki, Osaka 567-0047, Japan \\
$^2$Vin\v ca Institute for Nuclear Sciences (Physics Lab
010), P.O.Box 522, 11001 Belgrade, Serbia \\
$^3$Department of Physics and State Key Laboratory of Nuclear
Physics and Technology Peking University, Beijing 100871, China}


\date{\today}

\begin{abstract}
We study the $SU(3)$ linear sigma model for the pseudoscalar mesons
in the Gaussian Functional Approximation (GFA). We use the $SU(3)$
linear sigma model Lagrangian with nonet scalar and pseudo-scalar
mesons including symmetry breaking terms. In the GFA, we take the
Gaussian Ansatz for the ground state wave function and apply the
variational method to minimize the ground state energy. We derive
the gap equations for the dressed meson masses, which are actually
just variational parameters in the GFA method. We use the
Bethe-Salpeter equation for meson-meson scattering which provides
the masses of the physical nonet mesons. We construct the projection
operators for the flavor $SU(3)$ in order to work out the scattering
T-matrix in an efficient way. In this paper, we discuss the
properties of the Nambu-Goldstone bosons in various limits of the
chiral $U_L(3)\times U_R(3)$ symmetry.
\end{abstract}
\keywords{$SU(3)$ Linear Sigma Model, Gaussian Functional
Approximation, Nambu-Goldstone theory, projection operator}
\pacs{13.75.Lb, 11.30.Qc, 11.10.Cd, 11.10.St}
\maketitle

\section{Introduction}

The masses and properties of the $SU(3)$ scalar mesons are long
standing puzzles in hadron-nuclear physics related to the underlying
chiral symmetry of QCD. It is also very interesting to describe the
properties of these $SU(3)$ scalar and pseudoscalar mesons at finite
temperature and density. To this end it is important to study the
$U_L(3)\times U_R(3)$ symmetric linear sigma model in the
non-perturbative Gaussian Functional
Approximation~\cite{Cornwall:1974vz,Barnes:1978cd}. The linear sigma
model is a strongly interacting renormalizable quantum field theory;
due to the size of the self-interaction coupling constant(s) the
perturbative approximations seem to be inapplicable. Therefore, a
non-perturbative approximation, such as the Gaussian functional one,
that is equivalent to the resummation of certain infinite classes of
Feynman diagrams~\cite{Cornwall:1974vz,Barnes:1978cd}, is called
for.

The chiral $U_L(3)\times U_R(3)$ symmetry in the $SU(3)$ linear
sigma model is both spontaneously and explicitly broken, which means
that some pseudoscalar mesons are Nambu-Goldstone (NG) bosons, i.e.
with vanishing masses in the chiral limit. Straightforward solutions
to the gap equations in the Gaussian wave functional approximation
yield non-zero meson masses even in the chiral
limit~\cite{Cornwall:1974vz,Barnes:1978cd}, however, the proof of
the NG theorem used to be an open problem for over 30
years~\cite{Cornwall:1974vz,Dmitrasinovic:2003yv}. The first
solution to this problem in the $O(2)$ symmetric sigma model was
based on the Bethe-Salpeter equation~\cite{Dmitrasinovic:1994bq},
but other proofs soon followed~\cite{Aouissat:1997nu}. The first
proof was straightforwardly extended to $O(4)\simeq SU_L(2)\times
SU_R(2)$ in Ref.~\cite{Nakamura:2002ry} and was finally proven in
the general $O(N)$ case in Ref.~\cite{Dmitrasinovic:2003yv}. The
$SU(3)$ linear sigma model corresponds to a subgroup of the broken
$O(18)$ symmetry, the specifics of which depend on the (symmetry
breaking) parameters of the model, and thus readily fit into this
framework, but the Nambu-Goldstone theorem has never been explicitly
verified in the various limits of the chiral $U_L(3)\times U_R(3)$
Lagrangian.

There are also several influential studies of the thermal properties
of various spinless mesons that are based on the Gaussian
approximation, both in the two-flavor
$SU(2)$~\cite{Roh:1996ek,Chiku:1997va,Chiku:1998kd} and the
three-flavor $SU(3)$ cases~\cite{Lenaghan:2000ey}, but again without
taking into account the Bethe-Salpeter equation. Therefore these
studies do not obey the NG theorem in the chiral limit and as such
are ill-suited for the study of chiral symmetry restoration.

As for the $SU(3)$ case, there are many studies of the properties of
$U(3)$ mesons in the mean field, or the Born
approximation~\cite{Crater:1970yf,Chan:1974rb} and in the Gaussian
approximation~\cite{Lenaghan:2000ey}. The extension of the chiral
$SU(2)$ model to the chiral $SU(3)$ model is not trivial, because
there are several different self-interaction terms (three rather
than one in the simplest $SU(2)$ case, but one of them, the
$\lambda_2$, is generally expected to be (much) smaller than the
first one $\lambda_1$~\cite{Dmitrasinovic:1996fi}), so we have to
develop the necessary mathematical tools to deal with the scattering
Bethe-Salpeter equation (T-matrix) for the $U(3)$
mesons~\cite{Rebbi:1970vd,Crater:1970yf,Chan:1974rb}.

It is important to explicitly work out the NG bosons for various
cases of the chiral Lagrangian, so as to verify which pseudoscalar
mesons are NG bosons, before applying this formalism to non-zero
temperature and/or density. The question of $U_A(1)$ symmetry
breaking also looms large over this endeavor, so we pay special
attention to the flavor singlet-octet mixing.

As this is a complicated method applied to a difficult problem, and
many missteps have been made in the past, we take a step-by-step
approach.  We look first at the chiral limit: even here there are
some non-trivial cases, such as when $\lambda_2 = 0$ and c=0, (many)
new naively unexpected Nambu-Goldstone bosons appear beyond the
``elementary fields'' that already exist in the Lagrangian - they
are ``composite'' (bound state) NG bosons that correspond to the
broken $O(18)$ symmetry rather than the $U_L(3)\times U_R(3)$ one
that has (at most) nine NG bosons. This formation of composite NG
bosons demonstrates the non-perturbative nature and the respect of
the underlying symmetries by the GFA method.  We then turn on
explicit chiral symmetry breaking term $h_0 \neq 0$, but with good
$SU(3)$ symmetry. We show how this explicit symmetry breaking term
influences the masses of the pseudoscalar mesons to lowest (linear)
approximation.

In this paper, we present the necessary mathematical expressions
necessary for the application of the Gaussian functional
approximation, defined in Sect.~\ref{s: gauss func app}, to the
$SU(3)$ linear sigma model introduced in Sect.~\ref{s: lin sig mod}.
In Sect.~\ref{s: T matrix} we provide the expressions for the
Bethe-Salpeter equations in various channels using the $SU(3)$
projection operators developed in Sect.~\ref{s: proj op}. In
Sects.~\ref{s: NG theor I} we verify explicitly the NG theorem for
various cases and identify which are the NG bosons. In
Sect.~\ref{sec:ng} we briefly discuss the role of explicit symmetry
breaking terms by taking the simplest case. Sect.~\ref{s:conclusion}
is devoted to a summary of this paper.

\section{The $SU(3)$ Linear Sigma Model}
\label{s: lin sig mod}

To understand the masses of scalar and pseudoscalar mesons, we
employ the $SU(3)$ linear sigma model and use the Gaussian
Functional Approximation (GFA). In this section, we briefly review
the $SU(3)$ linear sigma model and work out the mass gap equations
in the mean field approximation.

The Lagrangian density of the ${U}_L(3) \times {U}_R(3)$ linear
sigma model is given by
\begin{eqnarray} \label{eq:lag1}
\lag (\Phi) &=& \mbox{Tr}(\partial_\mu \Phi^\dagger \partial^\mu \Phi -
m^2 \Phi^\dagger \Phi) - \lambda_1 [\mbox{Tr}(\Phi^\dagger \Phi)]^2
- \lambda_2 \mbox{Tr}(\Phi^\dagger \Phi)^2  \nonumber \\
& & + c [ \mbox{Det}(\Phi) + \mbox{Det}(\Phi^\dagger)]  +
\mbox{Tr}[H(\Phi + \Phi^\dagger)]~.
\end{eqnarray}
The meson field matrix $\Phi$ is a complex 3$\times$3 matrix of
the scalar and pseudo-scalar meson nonets,
\begin{eqnarray}
\Phi = T_a \phi_a = T_a (\sigma_a + i \pi_a)~,
\end{eqnarray}
where $\sigma_a$ are the scalar fields and $\pi_a$ are the
pseudoscalar fields. $T_a = \lambda_a/2$ are the generators of
$U(3)$, where $\lambda_a$ are the Gell-Mann matrices with $\lambda_0
= \sqrt{\frac{2}{3}}{\bf 1}$. The 3$\times$3 matrix $H$ breaks the
chiral symmetry explicitly and is chosen as
\begin{eqnarray}
H = T_a h_a~,
\end{eqnarray}
where $h_a$ are nine (external) $SU(3)$ symmetry breaking
parameters. Only three (diagonal) ones, $a=(0,3,8)$, are relevant
and the two, $a=(0,8)$, are the dominant ones. In this paper, we
only study the case $h_0 \neq 0$, and so $SU(3)$ symmetry is
conserved. We need to know at least the order of magnitude of the
coupling constants. Here we may use Ref.~\cite{Lenaghan:2000ey}
results as a (rough) guide to the expected values of the coupling
constants: to first approximation we expect $\lambda_1 \simeq 50,
\lambda_2 \simeq 1.5$, and if we define $c = \lambda_3 f_{\pi}$, we
find $\lambda_3 \simeq 50$. Thus we see that this is indeed a
strongly coupled system and that we need a non-perturbative
approximation.


The generators of $U(3)$ satisfy the (anti)commutation relations:
\begin{eqnarray}
[\lambda_a,\lambda_b]= 2if_{abc}\lambda_c\, ,~~{\rm and }~~\,
\{\lambda_a,\lambda_b\}= 2d_{abc}\lambda_c \, .
\end{eqnarray}
where $d_{abc}$ and $f_{abc}$ ``structure constants" are defined to
contain the 0 index. The values of $SU(3)$ structure constants are
provided in any good textbook and in review
articles~\cite{Gasiorowicz:1969kn}. Those $f$ structure constants
with a zero among its $a,b,c$ indices are zero and those $d$
structure constants containing 0 in its $a,b,c$ indices are only
non-zero for $d_{0ab} = \sqrt{2/3} \delta_{ab}$ with $a,b=1, \ldots
, 8$.

By inserting the $\Phi$ field into the Lagrangian, the
following Lagrangian is obtained.
\begin{eqnarray}
\label{eq:lag2} \lag (\sigma_a, \pi_a) &=&
\frac 12 [\partial_\mu
\sigma_a \partial^\mu \sigma_a+
\partial_\mu \pi_a \partial^\mu \pi_a ]
- \frac 12 m^2 (\sigma_a \sigma_a + \pi_a \pi_a)  \nonumber \\
&+& \gcon_{abc} ( \sigma_a \sigma_b \sigma_c - 3 \pi_a \pi_b
\sigma_c )
- 2 \hh_{abcd} \sigma_a \sigma_b \pi_c \pi_d   \nonumber \\
& & - \frac 13 \ff_{abcd} (\sigma_a \sigma_b \sigma_c \sigma_d +
\pi_a \pi_b \pi_c \pi_d ) + h_a \sigma_a.
\end{eqnarray}
The coefficients $\gcon_{abc}$, $\ff_{abcd}$ and $\hh_{abcd}$ are
given by
\begin{eqnarray}
\gcon_{abc} &=& \frac c6 [d_{abc} - \frac 32 (\delta_{a0} d_{0bc} +
\delta_{b0} d_{a0c}
+ \delta_{c0} d_{ab0} ) 
+ \frac 92 d_{000} \delta_{a0} \delta_{b0} \delta_{c0} ]~,  \\
\ff_{abcd} &=& \frac {\lambda_1}{4} (\delta_{ab} \delta_{cd} +
\delta_{ad} \delta_{bc}
+ \delta_{ac} \delta_{bd} ) 
+ \frac{\lambda_2}{8} ( d_{abn} d_{ncd} + d_{adn} d_{nbc}
+ d_{acn} d_{nbd} )~, \\
\hh_{abcd} &=& \frac {\lambda_1}{4} \delta_{ab} \delta_{cd} +
\frac{\lambda_2}{8} ( d_{abn} d_{ncd} + f_{acn} f_{nbd} + f_{bcn}
f_{nad} )~.
\end{eqnarray}
Considering the shift of the vacuum expectation values of
$\sigma_a=\bar \sigma_a+\sigma'_a$, the Lagrangian can be written as
\begin{eqnarray}\label{eq:lag3}
\lag (\sigma_a, \pi_a) &=& \frac 12 [ \partial_\mu \sigma_a
\partial^\mu \sigma_a
+  \partial_\mu \pi_a \partial^\mu \pi_a - \sigma_a (m_S^2)_{ab}
\sigma_b
- \pi_a (m_P^2)_{ab} \pi_b ]  \nonumber \\
& & + (\gcon_{abc} - \frac 43 \ff_{abcd} \bar{\sigma_d}) \sigma_a
\sigma_b \sigma_c - 3(\gcon_{abc} + \frac 43 \hh_{abcd}
\bar{\sigma_d}) \pi_a \pi_b \sigma_c
- 2 \hh_{abcd} \sigma_a \sigma_b \pi_c \pi_d   \nonumber \\
& & - \frac 13 \ff_{abcd} (\sigma_a \sigma_b \sigma_c \sigma_d +
\pi_a \pi_b \pi_c \pi_d ) - U(\bar{\sigma})~,
\end{eqnarray}
where we have just written $\sigma_a$ instead of $\sigma'_a$ for
simplicity of writing. The potential term $U(\bar{\sigma})$ is the
tree-approximation potential and $\bar \sigma_a$ is determined at
the tree level. The tree-level potential is
\begin{eqnarray}
U(\bar
\sigma_a)= \frac{m^2}{2}\bar\sigma_a^2 - \gcon_{abc}\bar\sigma_a
\bar\sigma_b \bar\sigma_c +\frac{1}{3}\ff_{abcd}\bar\sigma_a
\bar\sigma_b \bar\sigma_d \bar\sigma_d-h_a\bar\sigma_a~.
\end{eqnarray}
The mean field $\bar \sigma_a$ is obtained by the
variation.
\begin{eqnarray}
\frac{\partial U(\bar \sigma_a)}{\partial
\bar\sigma_a}=m^2\bar\sigma_a - 3\gcon_{abc} \bar\sigma_b
\bar\sigma_c+\frac{4}{3}\ff_{abcd} \bar\sigma_b \bar\sigma_c
\bar\sigma_d-h_a=0 \, ,
\end{eqnarray}
and the ``tree level'' masses of the scalar and pseudoscalar mesons
are given by
\begin{eqnarray}
(m_S^2)_{ab} &=& m^2 \delta_{ab} - 6\gcon_{abc} \bar \sigma_c
+ 4\ff_{abcd} \bar{ \sigma_c} \bar{ \sigma_d}~, \nonumber \\
(m_P^2)_{ab} &=& m^2 \delta_{ab} + 6\gcon_{abc} \bar \sigma_c +
4\hh_{abcd} \bar{ \sigma_c} \bar{ \sigma_d}~.
\end{eqnarray}
In general $SU(3)$ symmetry breaking case, the $0$-$8$
off-diagonal components of these matrices are not zero. The physical
states must have a diagonal mass matrix, and we have to diagonalize
the mass matrices.



\section{Gaussian functional approximation}
\label{s: gauss func app}

The Gaussian functional approximation
(GFA)~\cite{Barnes:1978cd,Dmitrasinovic:1994bq,Nakamura:2002ry} is
the method based on assuming the ground state solution is a Gaussian
functional around the mean field. We get the effective potential by
acting the Hamiltonian on the ground state with  scalar mesons
having vacuum expectation values.

First the \sch  ~equation in the functional formalism is given as
\begin{eqnarray}
H|0\ket = E|0\ket~,
\end{eqnarray}
where $H$ is the total Hamiltonian and $E$ is the corresponding energy for the
wave function $|0\ket$. The effective potential can be obtained by
\begin{eqnarray}
E = \int d^3x ~\langle 0 | \ham | 0 \rangle ~,
\end{eqnarray}
where the Hamiltonian density is obtained through the Legendre
transformation as
\begin{eqnarray}\label{eq:ham1}
\ham (\sigma_a, \pi_a) &=& - \frac 12 \frac{\delta^2}{\delta
\sigma^2_a} + \frac 12 (\nabla \sigma_a)^2 - \frac 12
\frac{\delta^2}{\delta \pi^2_a} + \frac 12 (\nabla \pi_a)^2 + \frac
12 m^2 (\sigma_a^2 +\pi_a^2)
\nonumber \\
& & - \gcon_{abc} ( \sigma_a \sigma_b \sigma_c - 3 \pi_a \pi_b
\sigma_c)
+ 2 \hh_{abcd} \sigma_a \sigma_b \pi_c \pi_d   \nonumber \\
& & + \frac 13 \ff_{abcd} (\sigma_a \sigma_b \sigma_c \sigma_d +
\pi_a \pi_b \pi_c \pi_d ) - h_a \sigma_a~.
\end{eqnarray}
and the ground state wave functional as a Gaussian function is
\begin{eqnarray}
| 0 \rangle &=& N\exp [ -\frac 14 (\sigma_a - \bar \sigma_a )
G^{-1}_{ab}(m_\sigma) (\sigma_b - \bar \sigma_b) - \frac 14 \pi_a
G^{-1}_{ab}(m_\pi) \pi_b]~.
\end{eqnarray}
Here, $N$ is the normalization factor.  The mass propagator is written as
\begin{eqnarray}
G_{ab}(x,y)=\fra \delta_{ab}\int
\frac{d^3k}{(2\pi)^3}\frac{1}{\sqrt{\vec k^2 +m_a^2}}e^{i\vec k(\vec
x-\vec y)}~.
\end{eqnarray}
Finally the effective potential can be calculated as
\begin{eqnarray}
\varepsilon &=& \bra 0|\ham|0\ket\nn &=& \frac 12 m^2 \bar
\sigma_a^2 + \frac 14 \{
G^{-1}_{ab}(m_{\sigma}) + G^{-1}_{ab}(m_{\pi}) \} \nonumber \\
&+& \frac 12 (m^2 - m^2_{\sigma_a})G_{ab}(m_\sigma) + \frac 12 (m^2
- m^2_{\pi_a}) G_{ab}(m_\pi)
\nonumber \\
&-& \gcon_{abc} \{ \bar \sigma_a \bar \sigma_b \bar \sigma_c + 3
\bar \sigma_a ( G_{bc}(m_\sigma) - G_{bc}(m_\pi) ) \}
\nonumber \\
&+& 2 \hh_{abcd} \{ G_{ab}(m_\pi) \bar \sigma_c \bar \sigma_d +
G_{ab}(m_\pi) G_{cd}(m_\sigma) \} \nonumber \\
&+& \frac 13 \ff_{abcd} \{ \bar \sigma_a \bar \sigma_b \bar \sigma_c
\bar \sigma_d + 6\bar \sigma_a \bar \sigma_b G_{cd}(m_\sigma) \nonumber \\
&+&  3 G_{ab}(m_\sigma) G_{cd}(m_\sigma)
+ 3G_{ab}(m_\pi) G_{cd}(m_\pi) \} \nonumber \\
&-& h_a \bar {\sigma_a} \, .
\end{eqnarray}
The gap equations for scalar and pseudoscalar mesons are obtained
by applying the variational principle with respect to meson
masses, $\frac{\partial \varepsilon}{\partial m_a}=0$ and then the
masses are given as
\begin{eqnarray}\label{eq:gap1}
(m_S^2)_{ab} &=& m^2 \delta_{ab} - 6 \gcon_{abc} \bar \sigma_c + 4
\ff_{abcd} \bar \sigma_c \bar \sigma_d + 4 \ff_{abcd}
G_{cd}(m_\sigma) + 4 \hh_{abcd} G_{cd}(m_\pi) \, ,
\nonumber \\
(m_P^2)_{ab} &=& m^2 \delta_{ab} + 6 \gcon_{abc} \bar \sigma_c + 4
\hh_{abcd} \bar \sigma_c \bar \sigma_d + 4 \hh_{abcd} G_{cd}
(m_\sigma) + 4 \ff_{abcd} G_{cd}(m_\pi) \, .
\end{eqnarray}
The equation following from the variation of the energy density with respect to the mean field value $\frac{\partial \varepsilon}{\partial \bar \sigma_a} = 0$ is given by
\begin{eqnarray}\label{eq:gap2}
h_a &=& m^2 \bar \sigma_a - 3 \gcon_{abc} [ \bar \sigma_b \bar
\sigma_c + G_{bc}(m_\sigma) - G_{bc}(m_\pi) ] \nonumber \\
&+& 4\hh_{abcd}\bar \sigma_b G_{cd}(m_\pi) + \frac 13 \ff_{abcd} [ 4
\bar \sigma_b \bar \sigma_c \bar \sigma_d + 12 \bar \sigma_b G_{cd}
(m_\sigma) ] \, .
\end{eqnarray}
These equations provide the masses and the mean field values of the meson fields.

\section{T-matrix for meson-meson scattering}
\label{s: T matrix}

As explained in Refs.~\cite{Dmitrasinovic:2003yv,
Dmitrasinovic:1994bq} one should work out the T-matrix for the
determination of the pseudoscalar mesons in order to fulfill the NG
theorem. We work out the $\sigma$-$\pi$ scattering for pseudoscalar
mesons. The interaction kernel in the $\sigma$-$\pi$ channel is
written as
\begin{eqnarray}
-iV_{acbd}&=&-i2\hh_{acbd}\cdot 2 \cdot
2-i3(\gcon_{bea}+\frac{4}{3}\hh_{beaf} \bar\sigma_f)2
\frac{i}{s-m_e^2}(-i)3(\gcon_{edc} + \frac{4}{3}\hh_{edcg}\bar
\sigma_g)2\nn &=&-i[8\hh_{acbd} + 36(\gcon_{bea}+
\frac{4}{3}\hh_{beaf}\bar\sigma_f)\frac{1}{s -
m_e^2}(\gcon_{edc}+\frac{4}{3}\hh_{edcg}\bar\sigma_g)] \, .
\end{eqnarray}
With this interaction kernel we can get the T-matrix
as
\begin{eqnarray}
-iT_{abcd}&=&-iV_{abcd}-iV_{abef}i\Pi_{ef}(-iV_{efcd}) + ...
\\ \nonumber&=&-i(V_{abcd}+V_{abef}\Pi_{ef}T_{efcd}+...) \, .
\end{eqnarray}
Therefore, what we need to solve is the scattering matrix
\begin{eqnarray}\label{eq:Tmatrix}
T_{abcd}=V_{abcd}+V_{abef}\Pi_{ef} T_{efcd} \, .
\end{eqnarray}
The polarization term for meson masses $m_a$ and $m_b$ is
\begin{eqnarray}
i\Pi_{ab}(p^2)=\int \frac{i}{(k-p)^2-m_a^2+i\varepsilon}\frac{
i}{k^2-m_b^2+i\varepsilon}\frac{d^4k}{(2\pi)^4} \, .
\end{eqnarray}
To work out the polarization function $\Pi_{ab}$, first let us work
out the $p^2=0$ case.  In this case, we can write
\begin{eqnarray}\label{eq:Pi}
\Pi_{ab}(0) =
 i \int \left( \frac{1}{k^2 - m_a^2+i\varepsilon} -
\frac{1}{k^2-m_b^2+i\varepsilon} \right) \frac{1}{m_a^2-m_b^2}
\frac{d^4k}{(2\pi)^4} = {I_0(m_a^2)-I_0(m_b^2) \over
m_a^2-m_b^2} \, .
\end{eqnarray}
Here, we can write the integral as
\begin{eqnarray}
I_0(m^2)&=&i\int \frac{1}{k^2-m^2+i\varepsilon}
\frac{d^4k}{(2\pi)^4}= i \int \frac{1}{k_0^2-\vec
k^2-m^2+i\varepsilon} \frac{d^4k}{(2\pi)^4}\nn &=& i \int
\frac{1}{(k_0-\sqrt{\vec k^2+m^2}+i\varepsilon)(k_0+\sqrt{\vec
k^2+m^2}+i\varepsilon)} \frac{d^4k}{(2\pi)^4}\nn&=& \fra \int
\frac{d^3k}{(2\pi)^3}\frac{1}{\sqrt{\vec k^2+m^2}}=\frac{1}{4\pi^2}
\int_0^\Lambda \frac{k^2 dk}{\sqrt{k^2+m^2}}\nn&=&
\frac{1}{8\pi^2}m^2\left[ x_3\sqrt{1+x_3^2}-\log
|x_3+\sqrt{1+x_3^2}| \right] \, ,
\end{eqnarray}
where $x_3=\Lambda/m$. We take $\Lambda\sim 1$ GeV to be fixed as a
parameter of the model. We may take the 4 dimensional cut-off by
transforming the above integral as $k_0=ik_4$:
\begin{eqnarray}
I_0(m^2)&=&i\int \frac{1}{k^2-m^2+i\varepsilon}
\frac{d^4k}{(2\pi)^4}=\int \frac{1}{k^2+m^2}
\frac{d^4k}{(2\pi)^4}\nn&=& \frac{1}{2\pi^2}\int \frac{k^3
dk}{k^2+m^2}=\frac{1}{4\pi^2}m^2 \left[ x_4^2-\log |1+x_4^2| \right]
\, ,
\end{eqnarray}
where $x_4=\Lambda/m$.

We write here the case for $m_a=m_b=m$, which is written in the
paper of Nakamura et al.~\cite{Nakamura:2002ry}. The general case
has to be worked our in the same frame.
\begin{eqnarray}
\Pi_{aa}(s)=\Pi_{aa}(0)+\frac{s}{(4\pi)^2}(-1+J_{aa}(s)) \, ,
\end{eqnarray}
where
\begin{eqnarray}
J_{aa}(s)= \sqrt{ \frac{4m^2}{s}-1 } \arcsin \sqrt{ \frac{s}{4m^2} }
\, ,
\end{eqnarray}
for $\frac{s}{4m^2} <1 $, and
\begin{eqnarray}
J_{aa}(s)=\sqrt{1-\frac{4m^2}{s}}\left[\log
\left(\sqrt{\frac{s}{4m^2}}+\sqrt{\frac{s}{4m^2}-1}\right)-i\frac{\pi}{2}\right]\,
,
\end{eqnarray}
for $1<\frac{s}{4m^2}<\infty$. We should work out the general case
$m_a \ne m_b$, in the ``dispersive'' form (see Eqs. (\ref{e:eye1}))
\begin{eqnarray}
I_{M\mu}(s) &=& i \int {d^{4} k \over (2 \pi)^{4}} {1 \over
{\left[k^{2} - M^{2} + i \epsilon \right] \left[(k - P)^{2} -
\mu^{2} + i \epsilon \right]}}
\nonumber \\
&=& I_{M\mu}(0) - {s \over{(4 \pi)^2}} K_{M\mu}(s)
= {1\over{2 \lambda_{0}}} \left({\mu^{2} \over{\mu^{2} -
M^{2}}}\right)
 - {s \over{(4 \pi)^2}} K_{M\mu}(s)
\nonumber \\
&=& {1\over{2 \lambda_{0}}} \left({\mu^{2} \over{\mu^{2} -
M^{2}}}\right) - {s \over{16 \pi^3}} \int {d~t \over{t - s - i
\epsilon}} {\rm Im}~K_{M\mu}(t) ~,\ \label{e:eye2}
\end{eqnarray}
where $s = P^2$ and the real and imaginary parts are
\begin{eqnarray}
{\rm Im}~K_{M\mu}(s) &=& {1 \over s} {\rm Im}~I_{M\mu}(s)
\nonumber \\
&=& {\pi \over s} \sqrt{\left(1 - {(M - \mu)^2 \over s} \right)
\left(1 - {(M + \mu)^2 \over s} \right)} \theta(s - (M + \mu)^2) \,
,
\nonumber \\
{\rm Re}~K_{M\mu}(s) &=& {2 \over s} \Bigg[ \left({M^2 - \mu^2
\over 2s}\right) \log{M \over \mu} + {1 \over 2} \left( 1 +
\left({M^2 + \mu^2 \over M^2 - \mu^2} \right) \log{M \over \mu}
\right)
\nonumber \\
&-& \sqrt{\left(1 - {(M - \mu)^2 \over s} \right) \left(1 - {(M +
\mu)^2 \over s} \right)} {\rm tanh}^{-1} \sqrt{s - {(M + \mu)^2}
\over{s - {(M - \mu)^2} }} \Bigg]~. \ \label{e:eye1}
\end{eqnarray}

\section{$SU(3)$ Projection Operators and Mixing Operators}
\label{s: proj op}

To solve the scattering equation (\ref{eq:Tmatrix}), we use the
method of projection operators. The $SU(3)$ group structure for the
$SU(3)$ sigma model is:
\begin{eqnarray}\label{eq:flavor}
(\mathbf{1}_F \oplus \mathbf{8}_F) \otimes (\mathbf{1}_F \oplus
\mathbf{8}_F)&=&\mathbf{1}_{(1)} \oplus \mathbf{8}_{x(1)} \oplus
\mathbf{8}_{y(1)} \oplus
\\ \nonumber && \mathbf{1}_{(8)} \oplus \mathbf{8}_{S(8)} \oplus \mathbf{8}_{A(8)}
\oplus \mathbf{27}_{S(8)} \oplus \mathbf{10}_{A(8)} \oplus
\overline{\mathbf{10}}_{A(8)} \, ,
\end{eqnarray}
We can write out the corresponding projection operators with some
manipulations:
\begin{eqnarray}\label{eq:projnonmixing}
P^{1(1)}_{abcd} &=& \delta_{ab} \delta_{cd} {\Big |}_{a,b,c,d=0}  \,
,
\\ \nonumber P^{8x(1)}_{abcd} &=& {3\over2} \sum_{n=1}^8 \Big (d_{abn}  d_{cdn} \Big )
{\Big |}_{a,c=0,b,d=1\cdots8} \, ,
\\ \nonumber P^{8y(1)}_{abcd} &=& {3\over2} \sum_{n=1}^8 \Big ( d_{abn}  d_{cdn} \Big )
{\Big |}_{a,c=1\cdots8,b,d=0} \, ,
\\ \nonumber P^{1(8)}_{abcd} &=& {1\over8}
\delta_{ab} \delta_{cd} {\Big |}_{a,b,c,d=1\cdots8}  \, ,
\\ \nonumber P^{8S(8)}_{abcd} &=& {3\over5} \sum_{n=1}^8 \Big ( d_{abn}  d_{cdn} \Big
) {\Big |}_{a,b,c,d=1\cdots8}
\\ \nonumber P^{8A(8)}_{abcd} &=& {1\over3} \sum_{n=1}^8 \Big ( f_{abn}  f_{cdn} \Big
) \, ,
\\ \nonumber P^{27S(8)}_{abcd} &=& {1\over2} \Big ( \delta_{ac} \delta_{bd} +  \delta_{ad}
\delta_{bc}
\Big ) {\Big |}_{a,b,c,d=1\cdots8}  - P^{1(8)}_{abcd} -
P^{8S(8)}_{abcd}  \, ,\\
\nonumber P^{(10 + \overline{ 10})A(8)}_{abcd} &=& {1\over2} \Big ( \delta_{ac} \delta_{bd}
- \delta_{ad} \delta_{bc} \Big )  {\Big |}_{a,b,c,d=1\cdots8}  -
P^{8A(8)}_{abcd} \, .
\end{eqnarray}
where the indices $a,b,c,d$ can be $0\cdots8$. When they are 0, they
are related to $\mathbf{1}_F$ of Eq.~(\ref{eq:flavor}); while when they
are $1\cdots8$, they are related to $\mathbf{8}_F$ of
Eq.~(\ref{eq:flavor}). These projection operators satisfy
\begin{eqnarray}
P^x_{abef} P^y_{efcd} = \delta^{xy} P^x_{abcd} \, .
\end{eqnarray}

We have used various relations for the derivation of the projection
operators: \beq f_{aij}f_{bij}&=&3\delta_{ab} \, ,\nn
d_{aij}d_{bij}&=&\frac{5}{3}\delta_{ab}\, ,\nn d_{aij}f_{bij}&=&0\,
,\nn d_{aij}\delta_{ij}&=&0\, ,\nn d_{aij}d_{bjk}d_{cki}&=&-\fra
d_{abc}\, ,\nn f_{aij}f_{bjk}d_{cki}&=&-\frac{3}{2} d_{abc}\, ,\nn
f_{aij}f_{bjk}f_{cki}&=& \frac{2}{3} f_{abc}\, ,\nn
f_{aij}d_{bjk}d_{cki}&=& 0 \, ,\nn
d_{abn}d_{cdn}+d_{acn}d_{bdn}+d_{adn}d_{bcn}&=&\frac{1}{3}
(\delta_{ab}\delta_{cd}+\delta_{ac}\delta_{bd}+\delta_{ad}\delta_{bc})\,
,\nn
f_{acn}f_{bdn}+f_{adn}f_{bcn}-3d_{abn}d_{cdn}&=&-(\delta_{ac}\delta_{bd}
+\delta_{ad}\delta_{bc})+\delta_{ab}\delta_{cd}\, ,\nn
f_{acn}d_{bdn}-f_{adn}f_{bcn}-f_{abn}f_{cdn}&=& 0\, ,\nn
d_{acn}d_{bdn}-d_{adn}d_{bcn}-f_{abn}f_{cdn}&=& \frac{1}{12}
(\delta_{ad}\delta_{bc}-\delta_{ac}\delta_{bd})\, . \eeq

Besides these projection operators, we also have several mixing
operators. They are used to express the mixing between the singlet
and octet mesons. We note that they are not projection operators,
and so we use $O$ to denote them. In the singlet channel, there are
two operators:
\begin{eqnarray}\label{eq:projmix1}
O^{1(M1)}_{abcd} &=& \delta_{ab}\delta_{cd}{\Big
|}_{a,b=1\cdots8,c,d=0} \, ,
\\ \nonumber O^{1(M2)}_{abcd} &=& \delta_{ab}\delta_{cd}{\Big
|}_{a,b=0,c,d=1\cdots8} \, .
\end{eqnarray}
They provide the mixing of the singlet and octet mesons in the
resultant singlet channel. While in the octet channel, there are six
operators
\begin{eqnarray}\label{eq:projmix8}
O^{8(M1)}_{abcd} &=& \sum_{n=1}^8  d_{abn} d_{cdn}{\Big
|}_{a=0,b,c,d=1\cdots8} \, ,
\\ \nonumber
O^{8(M2)}_{abcd} &=&  \sum_{n=1}^8 d_{abn} d_{cdn}{\Big
|}_{a,b,d=1\cdots8,c=0} \, ,
\\ \nonumber O^{8(M3)}_{abcd} &=& \sum_{n=1}^8  d_{abn} d_{cdn}{\Big
|}_{b=0,a,c,d=1\cdots8} \, ,
\\ \nonumber O^{8(M4)}_{abcd} &=& \sum_{n=1}^8 d_{abn} d_{cdn}{\Big
|}_{a,b,c=1\cdots8,d=0} \, ,
\\ \nonumber O^{8(M5)}_{abcd} &=& \sum_{n=1}^8  d_{abn} d_{cdn}{\Big
|}_{a,d=0,b,c=1\cdots8} \, ,
\\ \nonumber O^{8(M6)}_{abcd} &=& \sum_{n=1}^8 d_{abn} d_{cdn}{\Big
|}_{b,c=0,a,d=1\cdots8} \, .
\end{eqnarray}
They provide the mixing of the singlet-octet mesons and the octet mesons
in the resultant octet channel.

We modify the scattering equation (\ref{eq:Tmatrix}) to be
\begin{eqnarray}\label{eq:T2}
T_{abcd} &=& V_{abcd} + V_{abef} \Pi_{ef} T_{efcd}
\\ \nonumber &=& V_{abcd} +
V_{aba^\prime b^\prime} \Pi_{a^\prime b^\prime c^\prime d^\prime}
T_{c^\prime d^\prime cd} \, ,
\end{eqnarray}
where $\Pi_{abcd}= \delta_{ac} \delta_{bd} \Pi_{ab}$. Then by using
the projection operators as well as the mixing operators, we find
that there is a unique expansion for $V_{abcd}$ of the scattering of
$\sigma \pi \rightarrow \sigma \pi$:
\begin{eqnarray}
V_{abcd} &=& V_{1(1)} P^{1(1)}_{abcd} + V_{8x(1)} P^{8x(1)}_{abcd} +
V_{8y(1)} P^{8y(1)}_{abcd} + \\ \nonumber && V_{1(8)}
P^{1(8)}_{abcd} + V_{8S(8)} P^{8S(8)}_{abcd} + V_{27S(8)}
P^{27S(8)}_{abcd} + V_{8A(8)} P^{8A(8)}_{abcd} + V_{10A(8)} P^{(10 +
\overline{ 10})A(8)}_{abcd} \\ \nonumber && + V_{1(M1)}
O^{1(M1)}_{abcd} + V_{1(M2)} O^{1(M2)}_{abcd} + V_{8(M1)}
O^{8(M1)}_{abcd} + V_{8(M2)} O^{8(M2)}_{abcd} \\ \nonumber && +
V_{8(M3)} O^{8(M3)}_{abcd} + V_{8(M4)} O^{8(M4)}_{abcd} + V_{8(M5)}
O^{8(M5)}_{abcd} + V_{8(M6)} O^{8(M6)}_{abcd} \, ,
\end{eqnarray}
and so does $\Pi_{abcd}$. Moreover, we find that for $\Pi_{abcd}$
only the projection operators are enough, which means:
\begin{eqnarray}
\Pi_{1(Mi)} = \Pi_{8(Mi)} = 0 \, .
\end{eqnarray}

Since we find that both $V_{abcd}$ and $\Pi_{abcd}$ can be expanded
by the projection operators (\ref{eq:projnonmixing}) and the mixing
operators (\ref{eq:projmix1}) and (\ref{eq:projmix8}), we assume
that the scattering matrix $T_{abcd}$ can also be expanded by these
operators:
\begin{eqnarray}
T_{abcd} &=& T_{1(1)} P^{1(1)}_{abcd} + T_{8x(1)} P^{8x(1)}_{abcd} +
T_{8y(1)} P^{8y(1)}_{abcd} + \\ \nonumber && T_{1(8)}
P^{1(8)}_{abcd} + T_{8S(8)} P^{8S(8)}_{abcd} + T_{27S(8)}
P^{27S(8)}_{abcd} + T_{8A(8)} P^{8A(8)}_{abcd} + T_{10A(8)} P^{(10 +
\overline{ 10})A(8)}_{abcd} \\ \nonumber && + T_{1(M1)}
O^{1(M1)}_{abcd} + T_{1(M2)} O^{1(M2)}_{abcd} + T_{8(M1)}
O^{8(M1)}_{abcd} + T_{8(M2)} O^{8(M2)}_{abcd} \\ \nonumber && +
T_{8(M3)} O^{8(M3)}_{abcd} + T_{8(M4)} O^{8(M4)}_{abcd} + T_{8(M5)}
O^{8(M5)}_{abcd} + T_{8(M6)} O^{8(M6)}_{abcd}\, ,
\end{eqnarray}
and then we can separate the scattering equation (\ref{eq:T2}) into
several equations in different channels. We discuss them in the
following subsections.

\subsection{Singlet Channel}

We have the following relations for all the operators:
\begin{eqnarray}
O^{1}_{abef} O^{x}_{efcd} = 0 \, ,
\end{eqnarray}
where $O^{1}$ denotes $P^{1(1)}$, $P^{1(8)}$, $O^{1(Mi)}$ and
$O^{1(M2)}$, and $O^{x}$ denotes operators of other flavors.
Therefore, we can write out the equation (\ref{eq:T2}) in the
singlet channels:
\begin{eqnarray}
&& T_{1(1)} P^{1(1)}_{abcd} + T_{1(8)} P^{1(8)}_{abcd} + T_{1(M1)}
O^{1(M1)}_{abcd} + T_{1(M2)} O^{1(M2)}_{abcd} \\ \nonumber &=& \Big
( V_{1(1)} P^{1(1)}_{abcd} + V_{1(8)} P^{1(8)}_{abcd} + V_{1(M1)}
O^{1(M1)}_{aba^\prime b^\prime} + V_{1(M2)} O^{1(M2)}_{aba^\prime
b^\prime} \Big ) + \\ \nonumber && \Big ( V_{1(1)}
P^{1(1)}_{aba^\prime b^\prime} + V_{1(8)} P^{1(8)}_{aba^\prime
b^\prime} + V_{1(M1)} O^{1(M1)}_{aba^\prime b^\prime} + V_{1(M2)}
O^{1(M2)}_{aba^\prime b^\prime} \Big ) \times \nn && \Big ( \Pi_{1(1)}
P^{1(1)}_{a^\prime b^\prime c^\prime d^\prime} + \Pi_{1(8)}
P^{1(8)}_{a^\prime b^\prime c^\prime d^\prime} \Big ) \times
\\ \nonumber && \Big ( T_{1(1)} P^{1(1)}_{c^\prime
d^\prime cd} + T_{1(8)} P^{1(8)}_{c^\prime d^\prime cd} + T_{1(M1)}
O^{1(M1)}_{c^\prime d^\prime cd} + T_{1(M2)} O^{1(M2)}_{c^\prime
d^\prime cd} \Big ) \, .
\end{eqnarray}
It can be simplified to:
\begin{eqnarray}\nonumber
&& \left ( \begin{array}{cc} T_{1(1)} & 2\sqrt2T_{1(M2)}
\\2\sqrt2T_{1(M1)} & T_{1(8)} \end{array} \right )
= \left ( \begin{array}{cc} V_{1(1)} & 2\sqrt2V_{1(M2)}
\\2\sqrt2V_{1(M1)} & V_{1(8)} \end{array} \right ) \\ \nonumber && +
\left ( \begin{array}{cc} V_{1(1)} & 2\sqrt2V_{1(M2)}
\\2\sqrt2V_{1(M1)} & V_{1(8)} \end{array} \right ) \left ( \begin{array}{cc} \Pi_{1(1)} & 0
\\0 & \Pi_{1(8)} \end{array} \right ) \left ( \begin{array}{cc} T_{1(1)} & 2\sqrt2T_{1(M2)}
\\2\sqrt2T_{1(M1)} & T_{1(8)} \end{array} \right ) \, ,
\end{eqnarray}
And its solution is
\begin{eqnarray}\nonumber\label{eq:solution1}
&& \left ( \begin{array}{cc} T_{1(1)} & 2\sqrt2T_{1(M2)}
\\2\sqrt2T_{1(M1)} & T_{1(8)} \end{array} \right )
\\ \nonumber &=& \left ( 1 - \left ( \begin{array}{cc} V_{1(1)} & 2\sqrt2V_{1(M2)}
\\2\sqrt2V_{1(M1)} & V_{1(8)} \end{array} \right ) \left ( \begin{array}{cc} \Pi_{1(1)} & 0
\\0 & \Pi_{1(8)} \end{array} \right ) \right )^{-1} \left ( \begin{array}{cc} V_{1(1)} &
2\sqrt2V_{1(M2)}
\\2\sqrt2V_{1(M1)} & V_{1(8)} \end{array} \right ) \, .
\end{eqnarray}

\subsection{Octet Channel ($\mathbf{8}_{x(1)}$, $\mathbf{8}_{y(1)}$ and $\mathbf{8}_{S(8)}$)}

We have the following relations for all the operators:
\begin{eqnarray}
O^{8}_{abef} O^{x}_{efcd} = 0 \, ,
\end{eqnarray}
where $O^{8}$ denotes $P^{8x(1)}$, $P^{8y(1)}$, $P^{8S(8)}$ and
$O^{8(Mi)}$, and $O^{x}$ denotes other operators. Therefore, we can
similarly write out the scattering equation (\ref{eq:T2}) in the
octet channels. After some simplifications it turns to be:
\begin{eqnarray}\nonumber
&& \left( \begin{array}{ccc} T_{8x(1)} & {2\over3}T_{8(M5)} &
{\sqrt{10}\over3} T_{8(M1)}
\\ {2\over3}T_{8(M6)} & T_{8y(1)} & {\sqrt{10}\over3} T_{8(M3)}
\\ {\sqrt{10}\over3} T_{8(M1)} & {\sqrt{10}\over3} T_{8(M4)} & T_{8S(8)}
\end{array}\right) = \left( \begin{array}{ccc} V_{8x(1)} & {2\over3}V_{8(M5)} &
{\sqrt{10}\over3} V_{8(M1)}
\\ {2\over3}V_{8(M6)} & V_{8y(1)} & {\sqrt{10}\over3} V_{8(M3)}
\\ {\sqrt{10}\over3} V_{8(M1)} & {\sqrt{10}\over3} V_{8(M4)} & V_{8S(8)}
\end{array}\right)+
\\ \nonumber &&  \left( \begin{array}{ccc} V_{8x(1)} & {2\over3}V_{8(M5)} &
{\sqrt{10}\over3} V_{8(M1)}
\\ {2\over3}V_{8(M6)} & V_{8y(1)} & {\sqrt{10}\over3} V_{8(M3)}
\\ {\sqrt{10}\over3} V_{8(M1)} & {\sqrt{10}\over3} V_{8(M4)} & V_{8S(8)}
\end{array}\right) \left( \begin{array}{ccc} \Pi_{8x(1)} & 0 &
0
\\ 0 & \Pi_{8y(1)} & 0
\\ 0 & 0 & \Pi_{8S(8)}
\end{array}\right) \left( \begin{array}{ccc} T_{8x(1)} & {2\over3}T_{8(M5)} &
{\sqrt{10}\over3} T_{8(M1)}
\\ {2\over3}T_{8(M6)} & T_{8y(1)} & {\sqrt{10}\over3} T_{8(M3)}
\\ {\sqrt{10}\over3} T_{8(M1)} & {\sqrt{10}\over3} T_{8(M4)} & T_{8S(8)}
\end{array}\right) \, .
\end{eqnarray}
And its solution is
\begin{eqnarray}\nonumber\label{eq:solution8}
&& \left( \begin{array}{ccc} T_{8x(1)} & {2\over3}T_{8(M5)} &
{\sqrt{10}\over3} T_{8(M1)}
\\ {2\over3}T_{8(M6)} & T_{8y(1)} & {\sqrt{10}\over3} T_{8(M3)}
\\ {\sqrt{10}\over3} T_{8(M1)} & {\sqrt{10}\over3} T_{8(M4)} & T_{8S(8)}
\end{array}\right) \\ \nonumber &=& \left ( 1 -
\left( \begin{array}{ccc} V_{8x(1)} & {2\over3}V_{8(M5)} &
{\sqrt{10}\over3} V_{8(M1)}
\\ {2\over3}V_{8(M6)} & V_{8y(1)} & {\sqrt{10}\over3} V_{8(M3)}
\\ {\sqrt{10}\over3} V_{8(M1)} & {\sqrt{10}\over3} V_{8(M4)} & V_{8S(8)}
\end{array}\right) \left( \begin{array}{ccc} \Pi_{8x(1)} & 0 &
0
\\ 0 & \Pi_{8y(1)} & 0
\\ 0 & 0 & \Pi_{8S(8)}
\end{array}\right) \right )^{-1} \times
\\ \nonumber && \left( \begin{array}{ccc} V_{8x(1)} & {2\over3}V_{8(M5)} &
{\sqrt{10}\over3} V_{8(M1)}
\\ {2\over3}V_{8(M6)} & V_{8y(1)} & {\sqrt{10}\over3} V_{8(M3)}
\\ {\sqrt{10}\over3} V_{8(M1)} & {\sqrt{10}\over3} V_{8(M4)} & V_{8S(8)}
\end{array}\right) \, .
\end{eqnarray}

\subsection{$\mathbf{8}_{A(8)}$, $(\mathbf{10}\oplus\overline{\mathbf{10}})_{A(8)}$ and $\mathbf{27}_{S(8)}$ Channels}

We have the following relations for all the operators:
\begin{eqnarray}
P^{8A(8)}_{abef} O^{x}_{efcd} = 0 \, ,
\end{eqnarray}
where $O^{x}$ denotes other operators. Therefore, we can write out
the scattering equation (\ref{eq:T2}) in the $\mathbf{8}_{A(8)}$
channel:
\begin{eqnarray}
T_{8A(8)} &=& V_{8A(8)}  + V_{8A(8)} \Pi_{8A(8)} T_{8A(8)} \, .
\end{eqnarray}

We have the following relations for all the operators:
\begin{eqnarray}
P^{(10+\overline{10})A(8)}_{abef} O^{x}_{efcd} = 0 \, ,
\end{eqnarray}
where $O^{x}$ denotes operators of other flavors. Therefore, we can
write out the scattering equation (\ref{eq:T2}) in the decuplet
channel:
\begin{eqnarray}
T_{10A(8)} &=& V_{10A(8)}  + V_{10A(8)} \Pi_{10A(8)} T_{10A(8)} \, .
\end{eqnarray}

We have the following relations for all the operators:
\begin{eqnarray}
P^{27S(8)}_{abef} O^{x}_{efcd} = 0 \, ,
\end{eqnarray}
where $O^{x}$ denotes operators of other flavors. Therefore, we can
write out the scattering equation (\ref{eq:T2}) in the
$\mathbf{27}_{S(8)}$ channel:
\begin{eqnarray}
T_{27S(8)} &=& V_{27S(8)}  + V_{27S(8)} \Pi_{27S(8)} T_{27S(8)} \, .
\end{eqnarray}

\section{The Nambu-Goldstone Theorem}
\label{s: NG theor I}

To check whether there are Nambu-Goldstone bosons, we need to
check whether the scattering matrix $T(s)$ has a pole at $s=0$. In
this section, we study the Nambu-Goldstone theorem, and verify the
Nambu-Goldstone bosons in the $SU(3)$ linear sigma model. We
assume that there is a $SU(3)$ symmetry, which means that only
$\bar \sigma_0$ is non-zero ($\bar \sigma_i = 0$, for
$i=1\cdots8$). To further simplify our calculation, we only study
the pseudoscalar channels ($\sigma \pi \rightarrow \sigma \pi$),
where the pseudoscalar mesons propagate in the interaction kernel.

Since the calculations in this system is still not so easy, our
analysis will be done step by step. First we assume $c=\lambda_2 =
0$, and $\lambda_1 \neq 0$. In this case, we find that all the
pseudoscalar mesons are Nambu-Goldstone bosons. Then we assume
$\lambda_2 \neq 0$, and find that only one singlet and one octet
pseudoscalar mesons remain Nambu-Goldstone bosons. Finally we assume
$c \neq 0$, which is the most general case conserving $SU(3)$
symmetry. We find that only one octet pseudoscalar mesons remain to
be Nambu-Goldstone bosons.

\subsection{Case I: $c=\lambda_2 = 0$}

This is the first step. When $c=\lambda_2 = 0$, we find
that there is no mixing between two flavor singlet mesons and
among the four flavor octet mesons in the pseudoscalar channel,
and we can expand the potential 
matrix $V$ ($\sigma$-$\pi$ scattering) by only using the
non-mixing projection operators (\ref{eq:projnonmixing}):
\begin{eqnarray}
V_{abcd} &=& V_{1(1)} P^{1(1)}_{abcd} + V_{8x(1)} P^{8x(1)}_{abcd} +
V_{8y(1)} P^{8y(1)}_{abcd} + \\ \nonumber && V_{1(8)}
P^{1(8)}_{abcd} + V_{8S(8)} P^{8S(8)}_{abcd} + V_{27S(8)}
P^{27S(8)}_{abcd} + V_{8A(8)} P^{8A(8)}_{abcd} + V_{10A(8)} P^{(10 +
\overline{ 10})A(8)}_{abcd} \, ,
\end{eqnarray}
where the coefficients $V_i$ are calculated to be
\begin{eqnarray}\label{eq:case1:vsp}
V_{1(1)}  &=& 2 \lambda_1 + {{ 4 \lambda_1^2 \bar \sigma_0^2 }\over{
s - (m_P^2)_{00} }} \, ,
\\ \nonumber V_{8x(1)} &=& 2 \lambda_1 + {{ 4 \lambda_1^2 \bar
\sigma_0^2 }\over{ s - (m_P^2)_{ii} }} \, ,
\\ \nonumber V_{8y(1)} &=& V_{1(8)} =
V_{8S(8)} = V_{8A(8)} = V_{27S(8)} = V_{10A(8)} = 2  \lambda_1 \, ,
\end{eqnarray}
To expand $\Pi_{ab}$ at the point $s=0$, first we write out its
four-index form following Eq.~(\ref{eq:Pi}):
\begin{eqnarray}
\Pi_{abcd}(s=0) = \delta_{ac} \delta_{bd} \Pi_{ab}(s=0) =
\delta_{ac} \delta_{bd} { I_0(m_S^{aa}) - I_0(m_P^{bb}) \over
(m_S^2)_{aa} - (m_P^2)_{bb}} \, ,
\end{eqnarray}
And $\Pi_{abcd}$ can also be expanded by using the projection
operators (\ref{eq:projnonmixing}):
\begin{eqnarray}
\Pi_{abcd} &=& \Pi_{1(1)} P^{1(1)}_{abcd} + \Pi_{8x(1)}
P^{8x(1)}_{abcd} + \Pi_{8y(1)} P^{8y(1)}_{abcd} + \\ \nonumber &&
\Pi_{1(8)} P^{1(8)}_{abcd} + \Pi_{8S(8)} P^{8S(8)}_{abcd} +
\Pi_{27S(8)} P^{27S(8)}_{abcd} + \Pi_{8A(8)} P^{8A(8)}_{abcd} +
\Pi_{10A(8)} P^{(10 + \overline{ 10})A(8)}_{abcd} \, ,
\end{eqnarray}
and its solution is:
\begin{eqnarray}\label{eq:case1:pi}
\Pi_{1(1)} &=& {I_0(m_S^{00}) - I_0(m_P^{00}) \over (m_S^2)_{00} -
(m_P^2)_{00}} \, ,
\\ \nonumber \Pi_{8x(1)} &=& {I_0(m_S^{00}) - I_0(m_P^{ii}) \over
(m_S^2)_{00} - (m_P^2)_{ii}} \, ,
\\ \nonumber \Pi_{8y(1)} &=& {I_0(m_S^{ii}) - I_0(m_P^{00}) \over
(m_S^2)_{ii} - (m_P^2)_{00}} \, ,
\\ \nonumber \Pi_{1(8)} &=& \Pi_{8S(8)} = \Pi_{8A(8)}
= \Pi_{27S(8)} = \Pi_{10A(8)} = {I_0(m_S^{ii}) - I_0(m_P^{ii}) \over
(m_S^2)_{ii} - (m_P^2)_{ii}} \, .
\end{eqnarray}

In the case of $c=0$ and $\lambda_2 = 0$, the mass
equations~(\ref{eq:gap1}) are diagonal,
so we have
\begin{eqnarray}\label{eq:case1:gap1}
(m_S^2)_{00} &=& m^2 + 3 \lambda_1 \bar \sigma_0^2  + 3 \lambda_1
I_0(m_S^{00}) + 8 \lambda_1 I_0(m_S^{ii})
 +  \lambda_1  I_0(m_P^{00}) +
8 \lambda_1  I_0(m_P^{ii}) \, , \\
\nonumber
(m_S^2)_{ii} &=& m^2 + \lambda_1 \bar \sigma_0^2+ \lambda_1
I_0(m_S^{00}) + 10 \lambda_1
I_0(m_S^{ii})  +  \lambda_1 I_0(m_P^{00}) +  8 \lambda_1
I_0(m_P^{ii}) \, ,
\end{eqnarray}
and
\begin{eqnarray}\label{eq:case1:gap2}
(m_P^2)_{00} &=& m^2 + \lambda_1 \bar \sigma_0^2 + \lambda_1
I_0(m_S^{00}) + 8 \lambda_1 I_0(m_S^{ii}) +  3\lambda_1
I_0(m_P^{00}) +  8 \lambda_1 I_0(m_P^{ii}) \, ,
\\ \nonumber (m_P^2)_{ii} &=& m^2 + \lambda_1 \bar \sigma_0^2  + \lambda_1
I_0(m_S^{00}) + 8 \lambda_1  I_0(m_S^{ii}) +  \lambda_1
I_0(m_P^{00}) + 10 \lambda_1 I_0(m_P^{ii}) \, ,
\end{eqnarray}
where $i = 1,\ldots, 8$, and $(m_S^2)_{ii}$ denote $(m_S^{11})^2$,
$(m_S^{22})^2$, etc.. Here we note that, as a matter of fact at
the tree approximation level, the octet scalar mesons, the singlet
pseudoscalar scalar meson and the octet pseudoscalar mesons all
have the same mass:
\begin{eqnarray}
(m_S^2)_{ii} = (m_P^2)_{00} = (m_P^2)_{ii} = m^2 + \lambda_1 \bar
\sigma_0^2 \, .
\end{eqnarray}
We note that one possible consistent solution to the gap Eqs.
(\ref{eq:gap2}) is that
they all (still) have the same mass:
\begin{eqnarray}\label{eq:case1:massrelation}
(m_S^2)_{ii} = (m_P^2)_{00} = (m_P^2)_{ii} = m^2 + \lambda_1 \bar
\sigma_0^2 + \lambda_1 I_0(m_S^{00}) + 19 \lambda_1  I_0(m_S^{ii})
\, .
\end{eqnarray}
This result is very interesting: whereas we expected to find
pseudo-scalar Nambu-Goldstone bosons, which have zero masses, we
found that even the pseudo-scalar mesons here have non-zero masses
$(m_P^2)_{00} = (m_P^2)_{ii}$. This is the usual ``problem'' of the
NG theorem in the Gaussian approximation.

From Eq.~(\ref{eq:gap2}), we have
\begin{eqnarray}\label{eq:case1:gap3}
m^2 &=& - \lambda_1 \bar \sigma_0^2 - 3 \lambda_1 I_0(m_S^{00}) - 8
\lambda_1 I_0(m_S^{ii}) - \lambda_1 I_0(m_P^{00}) -  8 \lambda_1
I_0(m_P^{ii}) \, .
\end{eqnarray}
Then using this equation together with Eqs.~(\ref{eq:case1:gap1})
and (\ref{eq:case1:gap2}), we obtain
\begin{eqnarray}\label{eq:case1:mass}
(m_S^2)_{00} &=& 2 \lambda_1 \bar \sigma_0^2 \, ,
\\ \nonumber (m_S^2)_{ii} &=& - 2 \lambda_1 I_0(m_S^{00}) + 2 \lambda_1
I_0(m_S^{ii}) \, ,
\\ \nonumber (m_P^2)_{00}&=& - 2 \lambda_1 I_0(m_S^{00}) + 2 \lambda_1
I_0(m_P^{00}) \, ,
\\ \nonumber  (m_P^2)_{ii} &=& - 2 \lambda_1 I_0(m_S^{00}) + 2 \lambda_1
I_0(m_P^{ii}) \, .
\end{eqnarray}

In order to check whether there are Nambu-Goldstone bosons, we
only need to check if the following equations hold
\begin{eqnarray}\label{eq:NGcheck}
V_i(s=0) \Pi_i(s=0) = 1 \, ,
\end{eqnarray}
because in this case there is no flavor mixing of T-matrix
elements/scattering operators. If Eq. (\ref{eq:NGcheck}) holds,
then the T-matrix elements subject to the following Bethe-Salpeter
equation
\begin{eqnarray}
T_i (s)= V_i (s) + V_i (s) \Pi_i (s) T_i (s) \, ,
\end{eqnarray}
have a pole at $s=0$. This means there are massless mesons
propagating, and thus the Nambu-Goldstone bosons turn up.

Eq.~(\ref{eq:NGcheck}) can be easily checked when $c=\lambda_2=0$.
By using Eqs.~(\ref{eq:case1:mass}) and (\ref{eq:case1:pi}), we have
\begin{eqnarray}
\Pi_{1(1)} &=& {(m_P^2)_{00} \over -4 \lambda_1 \bar \sigma_0^2 + 2
\lambda_1 (m_P^2)_{00}} \, ,
\\ \nonumber \Pi_{8x(1)} &=& {(m_P^2)_{ii} \over -4
\lambda_1 \bar \sigma_0^2 + 2 \lambda_1 (m_P^2)_{ii}} \, ,
\end{eqnarray}
and
\begin{eqnarray}
\Pi_{8y(1)} = \Pi_{1(8)} =  \Pi_{8S(8)} = \Pi_{8A(8)} = \Pi_{27S(8)}
= \Pi_{10A(8)} = {1\over2\lambda_1} \, ,
\end{eqnarray}
together with Eqs.~(\ref{eq:case1:vsp}) we have
\begin{eqnarray}
V_i(s=0) \Pi_i(s=0) = 1 \, ,
\end{eqnarray}
for all the allowed flavor representations ${\bf 1}_{(1)}, {\bf
8}_{x(1)}, {\bf 8}_{y(1)}, {\bf 1}_{(8)}, {\bf 8}_{S(8)}, {\bf
8}_{A(8)}, (\mathbf{10}\oplus\overline{\mathbf{10}})_{A(8)}, {\bf
27}_{S(8)}$.

As there are several flavor singlets and octets, it is not clear
just how many NG bosons in these channels are independent? Yet, it
is clear that there are at least 1+8+10+27=46 distinct NG bosons
when $c=0$ and $\lambda_2 = 0$. That is (much) more than 9 NG bosons
expected in the general $U_L(3)\times U_R(3)$ linear sigma model,
and more than 17 NG bosons when the $O(18)$ symmetry is broken down
to $O(17)$. The explanation for the fact that there are more than 17
NG bosons is that the $O(18)$ may be dynamically broken down to a
symmetry that is lower than $O(17)$, e.g. the $O(16)$ or even
$O(15)$. In this sense the GFA approximation is substantially
different from the Born, or the one-loop approximations, which are
not known to lead to ground state(s) with ``exotically broken''
symmetry.

Thus, we have proved that all the expected pseudoscalar mesons are
Nambu-Goldstone bosons when $c=0$ and $\lambda_2 = 0$, but also that
there are many more. As there are no ``elementary'' meson fields in
the flavor $(\mathbf{10}\oplus\overline{\mathbf{10}})_{A(8)}$ and
${\bf 27}_{S(8)}$-plets in the $SU(3)$ linear sigma model, we must
conclude that these NG bosons are (zero mass) bound states of
(massive) elementary boson fields. This goes to show that the GFA
method is well and truly non-perturbative and capable of dynamically
producing bound states even in exotic flavor channels, such as the
$(\mathbf{10}\oplus\overline{\mathbf{10}})_{A(8)}$ and ${\bf
27}_{S(8)}$. Of course, this does not mean that in the ground state
of QCD there are exotic NG bosons, because the $c=0$ and $\lambda_2
= 0$ conditions do not correspond to reality. Therefore we discuss
the $c= 0$ and $\lambda_2 \neq 0$ case next.

\subsection{Case II: $c= 0$ and $\lambda_2 \neq 0$}
\label{s: NG theor II}

When $c= 0$ and $\lambda_2 \neq 0$, the mixing between two singlet
pseudoscalar mesons and among three octet pseudoscalar mesons exist,
and we need to use the mixing operators. The scattering matrix $V$
can be expanded by using the non-mixing projection operators
(\ref{eq:projnonmixing}) as well as the mixing operators
(\ref{eq:projmix1}) and (\ref{eq:projmix8}), and the solution is
\begin{eqnarray}\label{eq:case1:Vabcd}
V_{1(1)} &=& 2 \lambda_1 + {2\over3} \lambda_2 +  {4\over9} {{ (3
\lambda_1 + \lambda_2)^2 \bar \sigma_0^2 }\over{ s - (m_P^2)_{00} }}
\, ,
\\ \nonumber V_{8x(1)} &=& 2 \lambda_1 + {2\over3} \lambda_2 +  {4\over9} {{ (3
\lambda_1 + \lambda_2)^2 \bar \sigma_0^2 }\over{ s - (m_P^2)_{ii}
}}\, ,
\\ \nonumber V_{8y(1)} &=& {2} \lambda_1 + {2\over3} \lambda_2 + {4\over9} {{
\lambda_2^2 \bar \sigma_0^2 }\over{ s - (m_P^2)_{ii} }}\, ,
\\ \nonumber V_{1(8)} &=&  2 \lambda_1 - {2\over3} \lambda_2 +  {32\over9} {{
\lambda_2^2 \bar \sigma_0^2 }\over{ s - (m_P^2)_{00} }}\, ,
\\ \nonumber V_{8S(8)} &=& 2 \lambda_1 - {4\over3} \lambda_2 +  {10\over9} {{
\lambda_2^2 \bar \sigma_0^2 }\over{ s - (m_P^2)_{ii} }}\, ,
\\ \nonumber V_{8A(8)} &=& 2  \lambda_1 + 6 \lambda_2\, ,
\\ \nonumber V_{27S(8)} &=& 2  \lambda_1 + 2 \lambda_2\, ,
\\ \nonumber V_{10A(8)} &=& 2  \lambda_1\, ,
\\ \nonumber V_{1(M1)} = V_{1(M2)} &=& {2\over3} \lambda_2 +  {4\over9} {{ (3
\lambda_1 \lambda_2 + \lambda_2^2) \bar \sigma_0^2 }\over{ s -
(m_P^2)_{00} }}\, ,
\\ \nonumber V_{8(M1)} = V_{8(M2)} &=& \lambda_2 +  {2\over3} {{ (3
\lambda_1 \lambda_2 + \lambda_2^2) \bar \sigma_0^2 }\over{ s -
(m_P^2)_{ii} }}\, ,
\\ \nonumber V_{8(M3)} = V_{8(M4)} &=& \lambda_2 +  {2\over3}
{{\lambda_2^2 \bar \sigma_0^2 }\over{ s - (m_P^2)_{ii}
}}\, ,
\\ \nonumber V_{8(M5)} = V_{8(M6)} &=& \lambda_2 +  {2\over3} {{ (3
\lambda_1 \lambda_2 + \lambda_2^2) \bar \sigma_0^2 }\over{ s -
(m_P^2)_{ii} }}\, .
\end{eqnarray}
We do the same procedure for $\Pi_{abcd}$, and the results are
(after choosing $s=0$)
\begin{eqnarray}\label{eq:case1:Piabcd}
\Pi_{1(1)} &=& {I_0(m_S^{00}) - I_0(m_P^{00}) \over (m_S^2)_{00} -
(m_P^2)_{00}}\, ,
\\ \nonumber \Pi_{8x(1)} &=& {I_0(m_S^{00}) - I_0(m_P^{ii}) \over (m_S^2)_{00}
- (m_P^2)_{ii}}\, ,
\\ \nonumber \Pi_{8y(1)} &=& {I_0(m_S^{ii}) - I_0(m_P^{00}) \over (m_S^2)_{ii}
- (m_P^2)_{00}}\, ,
\\ \nonumber \Pi_{1(8)} &=&  \Pi_{8S(8)} = \Pi_{8A(8)} =\Pi_{27S(8)} =\Pi_{10A(8)} ={I_0
(m_S^{ii}) - I_0(m_P^{ii}) \over (m_S^2)_{ii} - (m_P^2)_{ii}} \, ,
\\ \nonumber \Pi_{1(Mi)} &=& \Pi_{8(Mi)}
= 0\, .
\end{eqnarray}
So $\Pi_{abcd}$ is still ``diagonal''.

In the case of $c=0$ and $\lambda_2 \neq 0$, the masses are
diagonal, and from Eqs.~(\ref{eq:gap1}) we have
\begin{eqnarray}\label{eq:case2:gap1}
(m_S^2)_{00} &=& m^2 + 3 \lambda_1 \bar \sigma_0^2 + \lambda_2 \bar
\sigma_0^2  + (3 \lambda_1 + \lambda_2) I_0(m_S^{00}) + (8 \lambda_1
+ 8 \lambda_2) I_0(m_S^{ii}) \\ \nonumber && + ( \lambda_1 +
{1\over3}\lambda_2) I_0(m_P^{00}) + ( 8 \lambda_1 +
{8\over3}\lambda_2) I_0(m_P^{ii}) \, ,
\\ \nonumber (m_S^2)_{ii} &=& m^2 + \lambda_1 \bar \sigma_0^2 + \lambda_2 \bar
\sigma_0^2  + (\lambda_1 + \lambda_2) I_0(m_S^{00}) + (10 \lambda_1
+ 5 \lambda_2) I_0(m_S^{ii}) \\ \nonumber && + ( \lambda_1 +
{1\over3}\lambda_2) I_0(m_P^{00}) + ( 8 \lambda_1 +
{17\over3}\lambda_2) I_0(m_P^{ii}) \, ,
\end{eqnarray}
and
\begin{eqnarray}\label{eq:case2:gap2}
(m_P^2)_{00} &=& m^2 + \lambda_1 \bar \sigma_0^2 +
{1\over3}\lambda_2 \bar \sigma_0^2  + (\lambda_1 +
{1\over3}\lambda_2)
I_0(m_S^{00}) + (8 \lambda_1 + {8\over3} \lambda_2) I_0(m_S^{ii}) \\
\nonumber && + ( 3\lambda_1 + \lambda_2) I_0(m_P^{00}) + ( 8
\lambda_1 + 8\lambda_2) I_0(m_P^{ii})\, ,
\\ \nonumber (m_P^2)_{ii} &=& m^2 + \lambda_1 \bar \sigma_0^2 + {1\over3}\lambda_2 \bar
\sigma_0^2  + (\lambda_1 + {1\over3} \lambda_2) I_0(m_S^{00}) + (8
\lambda_1 + {17\over3} \lambda_2) I_0(m_S^{ii}) \\ \nonumber && + (
\lambda_1 + \lambda_2) I_0(m_P^{00}) + ( 10 \lambda_1 + 5\lambda_2)
I_0(m_P^{ii})\, .
\end{eqnarray}
From Eq.~(\ref{eq:gap2}), we have
\begin{eqnarray}\label{eq:case2:gap3}
m^2 &=& - \lambda_1 \bar \sigma_0^2 - {1\over3} \lambda_2 \bar
\sigma_0^2  - (3 \lambda_1 +  \lambda_2) I_0(m_S^{00}) - (8
\lambda_1 + 8 \lambda_2) I_0(m_S^{ii}) \\ \nonumber && - ( \lambda_1
+ {1\over3}\lambda_2) I_0(m_P^{00}) - ( 8 \lambda_1 +
{8\over3}\lambda_2) I_0(m_P^{ii}) \, .
\end{eqnarray}
Then using this equation together with Eqs.~(\ref{eq:case2:gap1})
and (\ref{eq:case2:gap2}), we obtain
\begin{eqnarray}\label{eq:case2:mass}
(m_S^2)_{00} &=& 2 \lambda_1 \bar \sigma_0^2 + {2\over3}\lambda_2
\bar \sigma_0^2 \, ,
\\ \nonumber (m_S^2)_{ii} &=& 2 \lambda_1 \big( I_0(m_S^{ii}) - I_0(m_S^{00}) \big)
+ {2\over3}\lambda_2 \bar
\sigma_0^2  +  3 \lambda_2\big( I_0(m_P^{ii}) - I_0(m_S^{ii})\big)
\, ,
\\ \nonumber (m_P^2)_{00} &=&  2\lambda_1 \big( I_0(m_P^{00}) -I_0(m_S^{00}) \big) +
{2\over3}\lambda_2 \big( I_0(m_P^{00}) - I_0(m_S^{00})\big) +
{16\over3}\lambda_2 \big(I_0(m_P^{ii}) - I_0(m_S^{ii}) \big) \, ,
\\ \nonumber (m_P^2)_{ii} &=&  2 \lambda_1\big( I_0(m_P^{ii}) -  I_0(m_S^{00})\big )
+ {2\over3} \lambda_2\big (I_0(m_P^{00}) - I_0(m_S^{00})\big)  +
{7\over3}\lambda_2\big ( I_0(m_P^{ii})  - I_0(m_S^{ii})\big ) \, .
\end{eqnarray}

\subsubsection{Singlet Channel}

Since the mixing exists, we need to use Eq.~(\ref{eq:solution1})
derived earlier. After inserting the expressions of the masses and
the polarization energies, $V_i$ and $\Pi_i$, which are listed in
Eqs.~(\ref{eq:case2:mass}), (\ref{eq:case1:Vabcd}) and
(\ref{eq:case1:Piabcd}), we can verify that at the kinematical
point $s=0$, we have
\begin{eqnarray}
\left | 1 - \left ( \begin{array}{cc} V_{1(1)} & 2\sqrt2V_{1(M2)}
\\2\sqrt2V_{1(M1)} & V_{1(8)} \end{array} \right ) \left ( \begin{array}{cc} \Pi_{1(1)} & 0
\\0 & \Pi_{1(8)} \end{array} \right ) \right | = 0
\, ,
\end{eqnarray}
with the meaning that the determinant of the matrix within the
vertical bars is zero. Therefore, there are Nambu-Goldstone
bosons. Since there is a mixing between the two singlet
pseudoscalar mesons, $T_{1(1)}$, $T_{1(8)}$, $T_{1(M1)}$ and
$T_{1(M2)}$ all have a pole at $s=0$. However, we can verify that
only one of the two eigenvalues is 0. This means that only one of
the singlet pseudoscalar meson is a Nambu-Goldstone boson.

\subsubsection{Octet Channel ($\mathbf{8}_{x(1)}$, $\mathbf{8}_{y(1)}$ and $\mathbf{8}_{S(8)}$)}

We calculate the solution Eq.~(\ref{eq:solution8}) when $c=0$ and
$\lambda_2 \neq 0$, and find that at the point $s=0$, we have
\begin{eqnarray}
\left | 1 - \left( \begin{array}{ccc} V_{8x(1)} & {2\over3}V_{8(M5)}
& {\sqrt{10}\over3} V_{8(M1)}
\\ {2\over3}V_{8(M6)} & V_{8y(1)} & {\sqrt{10}\over3} V_{8(M3)}
\\ {\sqrt{10}\over3} V_{8(M1)} & {\sqrt{10}\over3} V_{8(M4)} & V_{8S(8)}
\end{array}\right) \left( \begin{array}{ccc} \Pi_{8x(1)} & 0 &
0
\\ 0 & \Pi_{8y(1)} & 0
\\ 0 & 0 & \Pi_{8S(8)}
\end{array}\right) \right | = 0
\, ,
\end{eqnarray}
where, again the vertical bars denote the determinant of the
(3$\times$3) matrix within. Therefore, these equations describe
massless Nambu-Goldstone bosons. Moreover, we can verify that only
one of the three eigenvalues is zero. This is difficult to prove
analytically even with the aid of algebraic manipulation programs.
Therefore, we randomly choose the values for the relevant parameters
(coupling constants), and confirm this result. So we obtain the
result that only one octet and one singlet of pseudoscalar mesons
are Nambu-Goldstone bosons in this case. That agrees with the
conventional result in the Born approximation, although the
pseudoscalar spectral functions in the GFA
\cite{Dmitrasinovic:1998je} contain (much) more structure than a
simple Dirac delta function, see e.g. \cite{Nakamura:2002vn}.

\subsubsection{$\mathbf{8}_{A(8)}$, $(\mathbf{10}\oplus\overline{\mathbf{10}})_{A(8)}$ and $\mathbf{27}_{S(8)}$ Channels}

For the $\mathbf{8}_{A(8)}$ channel, we have
\begin{eqnarray}\nonumber
1 - V_{8A(8)} \Pi_{8A(8)} =  \lambda_2 {\bar \sigma_0^2 + (
I_0(m_S^{00})) - I_0(m_P^{00})) - 10 (I_0(m_S^{ii})-I_0(m_P^{ii}))
\over  (3\lambda_1 - \lambda_2)(I_0(m_S^{ii})-I_0(m_P^{ii})) +
\lambda_2(\bar \sigma_0^2 + I_0(m_S^{00})-I_0(m_P^{00}))} \, .
\end{eqnarray}
For the decuplet channel, we have
\begin{eqnarray}\nonumber
1 - V_{10A(8)} \Pi_{10A(8)} =  \lambda_2 {\bar \sigma_0^2 + (
I_0(m_S^{00})) - I_0(m_P^{00})) -  (I_0(m_S^{ii})-I_0(m_P^{ii}))
\over  (3\lambda_1 - \lambda_2)(I_0(m_S^{ii})-I_0(m_P^{ii})) +
\lambda_2(\bar \sigma_0^2 + I_0(m_S^{00})-I_0(m_P^{00}))} \, .
\end{eqnarray}
For the $\mathbf{27}_{S(8)}$ channel, we have
\begin{eqnarray}\nonumber
1 - V_{27S(8)} \Pi_{27S(8)} =  \lambda_2 {\bar \sigma_0^2 + (
I_0(m_S^{00})) - I_0(m_P^{00})) - 4 (I_0(m_S^{ii})-I_0(m_P^{ii}))
\over  (3\lambda_1 - \lambda_2)(I_0(m_S^{ii})-I_0(m_P^{ii})) +
\lambda_2(\bar \sigma_0^2 + I_0(m_S^{00})-I_0(m_P^{00}))} \, .
\end{eqnarray}
Therefore, none of them are Nambu-Goldstone bosons, as expected.

\subsection{Case III: $c \neq 0 \neq \lambda_2$}
\label{s: NG theor III}

In this case we assume that both $c$ and $\lambda_2$ are nonzero.
This is the most general case that conserves the $SU_L(3) \times
SU(3)_R$ symmetry. The scattering matrix $T$ can be expanded by
using the non-mixing projection operators (\ref{eq:projnonmixing})
as well as the mixing operators (\ref{eq:projmix1}) and
(\ref{eq:projmix8}), and its solution is
\begin{eqnarray}
V_{1(1)} &=& 2 \lambda_1 + {2\over3} \lambda_2 +  {2\over9} {{ 2(3
\lambda_1 + \lambda_2)^2 \bar \sigma_0^2 } + 2\sqrt{6} (3\lambda_1 +
\lambda_2) c \bar \sigma_0 + {3}c^2 \over{ s - (m_P^2)_{00} }} \, ,
\\ \nonumber V_{8x(1)} &=& 2 \lambda_1 + {2\over3} \lambda_2 +  {1\over18} {{8 (3
\lambda_1 + \lambda_2)^2 \bar \sigma_0^2 - 4 \sqrt6 (3\lambda_1 +
\lambda_2) c \bar \sigma_0 + 3 c^2 }\over{ s - (m_P^2)_{ii} }} \, ,
\\ \nonumber V_{8y(1)} &=& {2} \lambda_1 + {2\over3} \lambda_2 + {1\over18} {{
8\lambda_2^2 \bar \sigma_0^2 - 4 \sqrt6 \lambda_2 c \bar \sigma_0 +
3 c^2 }\over{ s - (m_P^2)_{ii} }} \, ,
\\ \nonumber V_{1(8)} &=&  2 \lambda_1 - {2\over3} \lambda_2 +  {4\over9}
{{8 \lambda_2^2 \bar \sigma_0^2- 4 \sqrt{6} \lambda_2 c \bar
\sigma_0 + 3 c^2 }\over{ s - (m_P^2)_{00} }} \, ,
\\ \nonumber V_{8S(8)} &=& 2 \lambda_1 - {4\over3} \lambda_2 +  {5\over9} {{
2\lambda_2^2 \bar \sigma_0^2 + 2 \sqrt6 \lambda_2 c \bar \sigma_0 +
3 c^2 }\over{ s - (m_P^2)_{ii} }} \, ,
\\ \nonumber V_{8A(8)} &=& 2  \lambda_1 + 6 \lambda_2 \, ,
\\ \nonumber V_{27S(8)} &=& 2  \lambda_1 + 2 \lambda_2 \, ,
\\ \nonumber V_{10A(8)} &=& 2  \lambda_1 \, ,
\\ \nonumber V_{1(M1)} = V_{1(M2)} &=& {2\over3} \lambda_2 +  {1\over9} {{ 4(3
\lambda_1 \lambda_2 + \lambda_2^2) \bar \sigma_0^2 - \sqrt6
(3\lambda_1 - \lambda_2) c \bar \sigma_0 - 3 c^2 }\over{ s -
(m_P^2)_{00} }} \, ,
\\ \nonumber V_{8(M1)} = V_{8(M2)} &=& \lambda_2 +  {1\over6} {{4 (3
\lambda_1 \lambda_2 + \lambda_2^2) \bar \sigma_0^2 +
\sqrt6(6\lambda_1 + \lambda_2) c \bar \sigma_0 - 3 c^2 }\over{ s -
(m_P^2)_{ii} }} \, ,
\\ \nonumber V_{8(M3)} = V_{8(M4)} &=& \lambda_2 +  {1\over6} {{4 \lambda_2^2
\bar \sigma_0^2 + \sqrt6 \lambda_2 c \bar \sigma_0  - 3 c^2}\over{ s
- (m_P^2)_{ii} }} \, ,
\\ \nonumber V_{8(M5)} = V_{8(M6)} &=& \lambda_2 +  {1\over12} {{ 8(3
\lambda_1 \lambda_2 + \lambda_2^2) \bar \sigma_0^2 - 2 \sqrt6
(3\lambda_1 + 2 \lambda_2) c \bar \sigma_0 + 3 c^2 }\over{ s -
(m_P^2)_{ii} }} \, .
\end{eqnarray}
We go through the same procedure for $\Pi_{ab}$, and the results
are (after setting $s=0$)
\begin{eqnarray}
\Pi_{1(1)} &=& {I_0(m_S^{00}) - I_0(m_P^{00}) \over (m_S^2)_{00} -
(m_P^2)_{00}} \, ,
\\ \nonumber \Pi_{8x(1)} &=& {I_0(m_S^{00}) - I_0(m_P^{ii}) \over (m_S^2)_{00}
- (m_P^2)_{ii}} \, ,
\\ \nonumber \Pi_{8y(1)} &=& {I_0(m_S^{ii}) - I_0(m_P^{00}) \over (m_S^2)_{ii}
- (m_P^2)_{00}} \, ,
\\ \nonumber \Pi_{1(8)} &=&  \Pi_{8S(8)} = \Pi_{8A(8)} =\Pi_{27S(8)} =\Pi_{10A(8)} ={I_0
(m_S^{ii}) - I_0(m_P^{ii}) \over (m_S^2)_{ii} - (m_P^2)_{ii}} \, ,
\\ \nonumber \Pi_{1(mix)} &=& \Pi_{8x(mix)} = \Pi_{8y(mix)} = \Pi_{8z(mix)}
= 0 \, ,
\end{eqnarray}
which is the same as the previous case.

In the case of $c \neq 0$ and $\lambda_2 \neq 0$, the masses are
still diagonal, and from Eqs.~(\ref{eq:gap1}) we have
\begin{eqnarray}\label{eq:case3:gap1}
(m_S^2)_{00} &=& m^2 + 3 \lambda_1 \bar \sigma_0^2 + \lambda_2 \bar
\sigma_0^2  + (3 \lambda_1 + \lambda_2) I_0(m_S^{00}) + (8 \lambda_1
+ 8 \lambda_2) I_0(m_S^{ii}) \\ \nonumber && + ( \lambda_1 +
{1\over3}\lambda_2) I_0(m_P^{00}) + ( 8 \lambda_1 +
{8\over3}\lambda_2) I_0(m_P^{ii}) - \sqrt{2\over3} c \bar \sigma_0
\, ,
\\ \nonumber (m_S^2)_{ii} &=& m^2 + \lambda_1 \bar \sigma_0^2 + \lambda_2 \bar
\sigma_0^2  + (\lambda_1 + \lambda_2) I_0(m_S^{00}) + (10 \lambda_1
+ 5 \lambda_2) I_0(m_S^{ii}) \\ \nonumber && + ( \lambda_1 +
{1\over3}\lambda_2) I_0(m_P^{00}) + ( 8 \lambda_1 +
{17\over3}\lambda_2) I_0(m_P^{ii}) +  \sqrt{1\over6} c \bar \sigma_0
\, ,
\end{eqnarray}
and
\begin{eqnarray}\label{eq:case3:gap2}
(m_P^2)_{00} &=& m^2 + \lambda_1 \bar \sigma_0^2 +
{1\over3}\lambda_2 \bar \sigma_0^2  + (\lambda_1 +
{1\over3}\lambda_2)
I_0(m_S^{00}) + (8 \lambda_1 + {8\over3} \lambda_2) I_0(m_S^{ii}) \\
\nonumber && + ( 3\lambda_1 + \lambda_2) I_0(m_P^{00}) + ( 8
\lambda_1 + 8\lambda_2) I_0(m_P^{ii}) + \sqrt{2\over3} c \bar
\sigma_0 \, ,
\\ \nonumber (m_P^2)_{ii} &=& m^2 + \lambda_1 \bar \sigma_0^2 + {1\over3}\lambda_2 \bar
\sigma_0^2  + (\lambda_1 + {1\over3} \lambda_2) I_0(m_S^{00}) + (8
\lambda_1 + {17\over3} \lambda_2) I_0(m_S^{ii}) \\ \nonumber && + (
\lambda_1 + \lambda_2) I_0(m_P^{00}) + ( 10 \lambda_1 + 5\lambda_2)
I_0(m_P^{ii}) - \sqrt{1\over6} c \bar \sigma_0 \, .
\end{eqnarray}
From Eq.~(\ref{eq:gap2}), we have
\begin{eqnarray}
m^2 &=& - \lambda_1 \bar \sigma_0^2 - {1\over3} \lambda_2 \bar
\sigma_0^2 + \sqrt{1\over6} c \bar \sigma_0  - (3 \lambda_1 +
\lambda_2)
I_0(m_S^{00}) - (8 \lambda_1 + 8 \lambda_2) I_0(m_S^{ii}) \\
\nonumber && - ( \lambda_1 + {1\over3}\lambda_2) I_0(m_P^{00}) - ( 8
\lambda_1 + {8\over3}\lambda_2) I_0(m_P^{ii})
\\ \nonumber && + \sqrt{1\over6} {c\over\bar\sigma_0} ( I_0(m_S^{00}) - I_0(m_P^{00}))
- 2 \sqrt{2\over3} {c\over\bar\sigma_0} ( I_0(m_S^{ii}) -
I_0(m_P^{ii})) \, ,
\end{eqnarray}
Therefore, by using this equation together with
Eqs.~(\ref{eq:case3:gap1}) and (\ref{eq:case3:gap2}), we obtain
\begin{eqnarray}\label{eq:case3:mass}
\nonumber (m_S^2)_{00} &=& 2 \lambda_1 \bar \sigma_0^2 +
{2\over3}\lambda_2 \bar \sigma_0^2 - \sqrt{1\over6} c \bar \sigma_0
+ \sqrt{1\over6} {c\over\bar\sigma_0} ( I_0(m_S^{00}) -
I_0(m_P^{00})) - 2 \sqrt{2\over3} {c\over\bar\sigma_0} (
I_0(m_S^{ii}) - I_0(m_P^{ii})) \, ,
\\ \nonumber (m_S^2)_{ii} &=& 2 \lambda_1 \big( I_0(m_S^{ii}) - I_0(m_S^{00}) \big)
+ \sqrt{2\over3} c \bar \sigma_0 + {2\over3}\lambda_2 \bar
\sigma_0^2  +  3 \lambda_2\big( I_0(m_P^{ii}) - I_0(m_S^{ii})\big)
\\ \nonumber && + \sqrt{1\over6} {c\over\bar\sigma_0} (
I_0(m_S^{00}) - I_0(m_P^{00})) - 2 \sqrt{2\over3}
{c\over\bar\sigma_0} ( I_0(m_S^{ii}) - I_0(m_P^{ii})) \, ,\\
\nonumber (m_P^2)_{00} &=&  2\lambda_1 \big( I_0(m_P^{00})
-I_0(m_S^{00}) \big) + \sqrt{3\over2} c \bar \sigma_0 +
{2\over3}\lambda_2 \big( I_0(m_P^{00}) - I_0(m_S^{00})\big) +
{16\over3}\lambda_2 \big(I_0(m_P^{ii}) - I_0(m_S^{ii}) \big) \\
\nonumber && + \sqrt{1\over6} {c\over\bar\sigma_0} ( I_0(m_S^{00})
- I_0(m_P^{00})) - 2 \sqrt{2\over3} {c\over\bar\sigma_0} (
I_0(m_S^{ii}) - I_0(m_P^{ii})) \, , \\
\nonumber (m_P^2)_{ii} &=&
2 \lambda_1\big( I_0(m_P^{ii}) -  I_0(m_S^{00})\big ) + {2\over3}
\lambda_2\big (I_0(m_P^{00}) - I_0(m_S^{00})\big)  +
{7\over3}\lambda_2\big ( I_0(m_P^{ii})  - I_0(m_S^{ii})\big )
\\ \nonumber && + \sqrt{1\over6} {c\over\bar\sigma_0} ( I_0(m_S^{00}) - I_0(m_P^{00}))
- 2 \sqrt{2\over3} {c\over\bar\sigma_0} ( I_0(m_S^{ii}) -
I_0(m_P^{ii})) \, .
\end{eqnarray}

\subsubsection{Singlet Channel}

We calculate the solution Eq.~(\ref{eq:solution1}) when $c\neq 0$
and $\lambda_2 \neq 0$, and find that at the point $s=0$, we have
\begin{eqnarray}
\left | 1 - \left ( \begin{array}{cc} V_{1(1)} & 2\sqrt2V_{1(M2)}
\\2\sqrt2V_{1(M1)} & V_{1(8)} \end{array} \right ) \left ( \begin{array}{cc} \Pi_{1(1)} & 0
\\0 & \Pi_{1(8)} \end{array} \right ) \right | \neq 0
\, ,
\end{eqnarray}
where, again the vertical bars denote the determinant of the
(2$\times$2) matrix within. Therefore, we have verified that the
singlet pseudoscalar meson is not a Nambu-Goldstone boson any more,
due to the $U_A(1)$ symmetry breaking interaction constant $c \neq
0$.

\subsubsection{Octet Channel ($\mathbf{8}_{x(1)}$, $\mathbf{8}_{y(1)}$ and $\mathbf{8}_{S(8)}$)}

We calculate the solution Eq.~(\ref{eq:solution8}) when $c\neq 0$
and $\lambda_2 \neq 0$, and find that at the point $s=0$, we have
\begin{eqnarray}
\left | 1 - \left( \begin{array}{ccc} V_{8x(1)} & {2\over3}V_{8(M5)}
& {\sqrt{10}\over3} V_{8(M1)}
\\ {2\over3}V_{8(M6)} & V_{8y(1)} & {\sqrt{10}\over3} V_{8(M3)}
\\ {\sqrt{10}\over3} V_{8(M1)} & {\sqrt{10}\over3} V_{8(M4)} & V_{8S(8)}
\end{array}\right) \left( \begin{array}{ccc} \Pi_{8x(1)} & 0 &
0
\\ 0 & \Pi_{8y(1)} & 0
\\ 0 & 0 & \Pi_{8S(8)}
\end{array}\right) \right | = 0
\, ,
\end{eqnarray}
where, again the vertical bars denote the determinant of the
(3$\times$3) matrix within. This is again difficult to prove
analytically, so we again randomly choose the values for the
relevant parameters, and obtain this result. Moreover, we can verify
that only one of its three eigenvalues is zero. So we obtain the
expected, yet non-trivial result that only one octet of pseudoscalar
mesons are Nambu-Goldstone bosons.

\section{Explicitly broken chiral symmetry and Dashen's formula}
\label{sec:ng}

When the chiral symmetry is explicitly broken, the NG theorem turns
into a relation between the chiral symmetry breaking parameter and
the NG boson mass, as first discussed by
Dashen~\cite{Dashen:1969eg,Dashen:1969ez}. The NG theorem in the
chiral limit has already been addressed in the Gaussian
approximation and equivalent formalisms in Refs.
\cite{Dmitrasinovic:1994bq,Aouissat:1997nu}. Here, we turn to the
non-chiral case.

As shown in Ref.~\cite{Dmitrasinovic:1994bq} in the chiral limit
the Nambu-Goldstone particle appears as a zero-mass pole in the
T-matrix in the pseudo-scalar channel. Next we look at the zero CM
energy $s = 0$ polarization function $V_{\pi}(0) \Pi_{\pi}(0) $ in
the non-chiral case $h_0 = \varepsilon \neq 0$. For simplicity's
sake we only study this in the $\lambda_2 = c = 0$ case (so as not
to have to deal with complications associated with channel
mixing(s) in the flavor-singlet and octet channels). Now the gap
Eq. (\ref{eq:gap2}) becomes
\begin{eqnarray}\label{eq:case4:gap3}
m^2 &=& {\epsilon \over \bar \sigma_0} - \lambda_1 \bar \sigma_0^2 -
3 \lambda_1 I_0(m_S^{00}) - 8 \lambda_1 I_0(m_S^{ii}) - \lambda_1
I_0(m_P^{00}) -  8 \lambda_1 I_0(m_P^{ii}) \, ,
\end{eqnarray}
together with (\ref{eq:case1:gap1}) and (\ref{eq:case1:gap2}),
we obtain
\begin{eqnarray}\label{eq:case4:mass}
(m_S^2)_{00} &=& {\epsilon \over \bar \sigma_0} + 2 \lambda_1 \bar
\sigma_0^2 \, ,
\\ \nonumber (m_S^2)_{ii} &=& {\epsilon \over \bar \sigma_0} -
2 \lambda_1 I_0(m_S^{00}) + 2 \lambda_1
I_0(m_S^{ii}) \, ,
\\ \nonumber (m_P^2)_{00} &=& {\epsilon \over \bar \sigma_0}
- 2 \lambda_1 I_0(m_S^{00}) + 2 \lambda_1 I_0(m_P^{00}) \, ,
\\ \nonumber (m_P^2)_{ii} &=& {\epsilon \over \bar \sigma_0}
- 2 \lambda_1 I_0(m_S^{00}) + 2 \lambda_1 I_0(m_P^{ii}) \, .
\end{eqnarray}

We work out the BS equation for the flavor-singlet and octet
channels. The polarization function is worked out in the
flavor-singlet channel $\mathbf{1}_{(1)}$ as
\begin{eqnarray}
V_{1(1)}(0) \Pi_{1(1)}(0) &=& 1 - {\epsilon \over \bar \sigma_0}
{(m_S^2)_{00} \over (m_P^2)_{00} ( (m_S^2)_{00} - (m_P^2)_{00} )} +
\mathcal{O}(\epsilon^2) \, ,
\end{eqnarray}
and we obtain the similar result for the flavor-octet channel
$\mathbf{8}_{x(1)}$:
\begin{eqnarray}
V_{8x(1)}(0) \Pi_{8x(1)}(0) &=& 1 - {\epsilon \over \bar \sigma_0}
{(m_S^2)_{00} \over (m_P^2)_{ii} ( (m_S^2)_{00} - (m_P^2)_{ii} )} +
\mathcal{O}(\epsilon^2) \, ,
\end{eqnarray}
as well as the positive-parity channel flavor-octet consisting of
two scalar mesons. Since $V_{\pi}(0) \Pi_{\pi}(0) \simeq 1 -
\mu^{-2} {\varepsilon \over v}$, we see that the pole in the
$s-$channel propagator has moved away from zero momentum. In order
to find the mass, we must take into account the residue at the pole;
thus we find $m_{\pi}^{2} = {\varepsilon \over v} + {\cal
O}(\varepsilon^{2})$, just as in the Born approximation. Here $h_0 =
\varepsilon$ is the explicit symmetry breaking parameter and the
``pion" symbol $\pi$ denotes the complete set of 17 pseudo-NG
bosons, whose mass should be small if this linearized approximation
is to hold. This result is valid only for ``small" values of the
explicit chiral symmetry breaking parameters, such as that of the
$SU(2)_L \times SU_R(2)$ symmetry breaking that is responsible for
the pion's mass.

Of course, with $h_0 \neq 0$ it is possible to have an explicit
breaking of the $O(18)$ symmetry down to the explicitly conserved,
yet spontaneously broken $O(17)$ symmetry. In that case there will
remain several massless NG bosons. For instance, in the other
flavor-singlet channel $\mathbf{1}_{(8)}$ made up of flavor-octet
mesons, we have
\begin{eqnarray}
V_{1(8)}(0) \Pi_{1(8)}(0) &=& 1 \, ,
\end{eqnarray}
and the same result holds for the $\mathbf{8}_{y(1)}$,
$\mathbf{8}_{S(8)}$, $\mathbf{8}_{A(8)}$, $\mathbf{27}_{S(8)}$ and
$(\mathbf{10}\oplus\overline{\mathbf{10}})_{A(8)}$ channels. Of
course, once one turns on $\lambda_2 \neq 0$ and/or $c \neq 0$ all
of these NG bosons acquire masses, as derived in Sect. \ref{s: NG
theor II}. The NG mesons also acquire masses when one explicitly
breaks the $O(17)$, $SU(3)$ or $SU(2)$ symmetries, e.g. by including
$h_8 \neq 0$ and/or $h_3 \neq 0$. These masses can be evaluated by
means of Dashen's formula so long as the explicit symmetry breaking
is small, which is not the case for realistic values of $h_8 \neq 0;
c \neq 0$. Therefore, this result is practically useful only for the
(iso-triplet) pion masses, but not for the kaons and the $\eta$ and
$\eta^{'}$ mesons. This is perhaps as far as one can go using only
analytic methods.

The next step, to be taken in our next paper, will be to numerically
solve the gap and Bethe-Salpeter equations with an explicitly broken
$SU(3)$ symmetry, so as to reproduce the experimental pseudo-scalar
masses and their weak decay constants and thus to fix all of the
free parameters in this model in the Gaussian approximation.

\section{Conclusion}
\label{s:conclusion}

We have studied the Nambu-Goldstone (NG) theorem for the
pseudo-scalar mesons in the $U(3)_L \times U_R(3)$ linear sigma
model. We have constructed the ground state wave function in the
Gaussian Functional Approximation (GFA). At this level, all the
scalar mesons and the pseudo-scalar mesons acquire finite masses by
the minimal spontaneous symmetry breaking $\sigma_0 \neq 0$. Hence,
the NG theorem is not satisfied at the GFA level.

Hence, we have developed a method to work out the Bethe-Salpeter
equation for the scattering T-matrix of mesons. To this end, it was
important to work out the projection operators in order to separate
various $SU(3)$ channels in the T-matrix. We have then explicitly
worked out the BS equations for the pseudo-scalar mesons in the
general case of the $U(3)_L \times U_R(3)$ linear sigma model. Since
the verification of the NG theorem is quite complicated and tricky,
we have decided to work out the NG theorem step by step.

Firstly we studied the $\lambda_1 \neq 0$ and $\lambda_2 = c = 0$
case. In this case, we verified that the NG bosons appear in the
usual flavor-nonet channel, where the NG bosons are present at the
mean field, or the Born approximation level. Additionally, we have
found new composite NG bosons in certain other flavor channels that
correspond to the breaking of the extended $O(18)$ symmetry down to
a lower symmetry.

Then we studied the case with $\lambda_1 \neq 0$ and $\lambda_2 \neq
0$, but $c=0$. In this case, we found the usual flavor-nonet of NG
bosons. We then studied the case when all the coupling constant in
the Lagrangian are non-zero. In this case, we found only the
flavor-octet pseudo-scalar mesons as the NG bosons: the ninth
pseudo-scalar meson acquires a non-zero mass and thus is not an NG
boson any more. Of course, $c \neq 0$ corresponds to the explicit
U$_A$(1) symmetry breaking, that affects the $\eta$ and $\eta^{'}$
mesons, and is comparable with, or perhaps even larger than the
explicit breaking of the $SU(3)_L \times SU_R(3)$ symmetry.

We have discussed another simple case in order to examine how
low-mass pseudo-NG bosons emerge due to the explicit chiral symmetry
breaking: when the Lagrangian has just one small explicit chiral
symmetry breaking parameter $h_0 = \varepsilon \neq 0$. There we
confirmed that Dashen's result for pseudo-NG boson masses hold in
the Gaussian approximation. This result is valid only for ``small"
values of the explicit chiral symmetry breaking parameters, such as
that of the $SU(2)_L \times SU_R(2)$ symmetry breaking that is
responsible for the pion's mass.

In this paper, we have analyzed the appearance of NG bosons for
various cases in the chiral $U(3)_L \times U_R(3)$ linear sigma
model Lagrangian. We have also studied the effect of the ``small"
explicit chiral symmetry breaking term to provide a small mass to
the pseudo-scalar bosons. This result is practically useful only for
the (iso-triplet of) pions, but not for the kaons and the $\eta$ and
$\eta^{'}$ mesons.

This is perhaps as far as one can possibly go using only analytic
methods. The next step, to be taken in our next paper, will be to
numerically solve the gap and Bethe-Salpeter equations with an
explicitly broken $SU(3)$ symmetry, so as to reproduce the
experimental pseudo-scalar masses and their weak decay constants and
thus to fix all of the free parameters in this model in the Gaussian
approximation. Then it will be possible and (very) interesting to
calculate the spectra of scalar bosons in the $SU(3)$ chiral sigma
model.

\section*{Acknowledgments}

One of us (H-X. C.) is grateful to the theory group of RCNP at Osaka
University for the four months stay as a Researcher and for fruitful
discussions with the theory members. Another of us (V.D.) wishes to
thank the Yukawa Institute for Theoretical Physics, Kyoto, for kind
hospitality at the YITP workshop ``New Frontiers in QCD 2010 -
Exotic Hadron Systems and Dense Matter'' (NFQCD10), where some of
this work was done. The work of H.T. is supported by JSPS under the
grant 70163962.

\end{document}